\documentclass{raa_twocolumn}
\usepackage{graphicx,times}
\usepackage{natbib}
\usepackage{amssymb,amsmath}
\bibpunct{(}{)}{;}{a}{}{,}

\usepackage{CJKutf8}

\usepackage[pagebackref=true]{hyperref}

\begin{document}
\begin{CJK*}{UTF8}{gbsn}

   \title{Overview of Ground-based Wide-Angle Cameras array}

 \volnopage{ {\bf 20XX} Vol.\ {\bf X} No. {\bf XX}, 000--000}
   \setcounter{page}{1}

   \author{Liping Xin\inst{1}, 
   Lei Huang\inst{1},
   Hongbo Cai\inst{1},
   Xuhui Han\inst{1}, 
   Yang Xu\inst{1},
   Xiaomeng Lu\inst{1},
   Huali Li\inst{1},
   Jing Wang\inst{1},
   Yulei Qiu\inst{1},
   Chao Wu\inst{1,7},
   Ruosong Zhang\inst{1},
   Pinpin Zhang\inst{1},
   Yujie Xiao\inst{1},
   Guangwei Li\inst{1},
   Jingsong Deng\inst{1,7},
   Dawei Xu\inst{1,7},
   Linjun Wang\inst{1},
   Jinran Xu\inst{1},
   Yinuo Ma\inst{1,8,9},
   Yangtong Zheng\inst{1},
   Wenlong Dong\inst{1},
   Zhuheng Yao\inst{1},
   Enwei Liang\inst{2},
   Xianggao Wang\inst{2},
   Xiangyu Wang\inst{3,4},
   Zigao Dai\inst{4,5},
   Yuangui Yang\inst{6},
   Jianyan Wei\inst{1,7}
   }

   \institute{National Astronomical Observatories, Chinese Academy of Sciences, Beijing 100101, China. {\it xlp@nao.cas.cn, wjy@nao.cas.cn}\\
        \and
        Guangxi Key Laboratory for Relativistic Astrophysics, School of Physical Science and Technology, Guangxi University, Nanning 530004, China\\
        \and
            Key Laboratory of Modern Astronomy and Astrophysics (Nanjing University), Ministry of Education, Nanjing 210093, China\\
        \and
        Department of Astronomy, University of Science and Technology of China, Hefei 230026, China\\
        \and
        School of Astronomy and Space Science, Nanjing University, Nanjing 210093, China\\    
        \and
             School of Physics and Electronic Information, Huaibei Normal University, Huaibei 235000, China\\
      	\and
	  School of Astronomy and Space Science, University of Chinese Academy of Sciences, Beijing 101408, China\\
      	\and
	  Institute for Frontier in Astronomy and Astrophysics, Beijing Normal University, Beijing 102206, People's Republic of China\\
            	\and
	  School of Physics and Astronomy, Beijing Normal University, Beijing 100875, People's Republic of China\\
\vs \no
   {\small Received 20XX Month Day; accepted 20XX Month Day}
}

\abstract{
As one of the key ground-based facilities of the Chinese-French \textit{SVOM} mission, the main scientific objectives of the Ground-based Wide Angle Camera array (GWAC) are to detect prompt optical emission of gamma-ray bursts or other short duration astronomical transients on a second-scale temporal resolution. 
GWAC is located at Xinglong observatory, China, and consists of  10 mounts and 40 cameras,  providing a joint field of view of about 3600 square degrees. 
The detection ability is 16 magnitude in 10 seconds of exposure time in the visual band under the condition of the new moon phase. 
Here, we give an overview of GWAC  and introduce the science motivation of the project, as well as the performance of the hardware and the software. The observation strategies and the data processing are briefly presented. The early sciences in the last 5 years since the first light are summarized. 
}

   \authorrunning{Xin et al. }            
   \titlerunning{}  
   \maketitle

%
\section{Introduction}           
\label{sect:intro}

The developments of deep, wide-field surveys in the past decade, such as the Zwicky Transient Facility (ZTF; \cite{2019PASP..131a8002B}),
the All-Sky Automated Survey for Supernovae (ASASSN; \cite{2014AAS...22323603S}), MASTER-Net project (\cite{2010AdAst2010E..30L}), the Asteroid Terrestrial-Impact Last Alert System (ATLAS; \cite{2011PASP..123...58T}),  the Gravitational-wave Optical Transient Observer (GOTO; \cite{2020MNRAS.497..726G}), The Large Array Survey Telescope (LAST; \cite{2023PASP..135f5001O})， 
the Evryscope (\cite{2023ApJS..265...63C}),  
the Wide Field Survey Telescope (WFST; \cite{2023SCPMA..6609512W}),
the Multi-channel Photometric Survey telescope (Mephisto; \cite{2020SPIE11445E..7MY})
and The Mini-SiTian Array or its pathfinder (\cite{2025RAA....25d4001H})
have led to the discoveries of a large sample of cosmic transients with  time scales from days to weeks, including supernovae, tidal disruption events and some fast blue optical transients (FBOTs).
The coming optical survey, the Rubin Observatory Legacy Survey of Space and Time (LSST, \cite{2019ApJ...873..111I}, would generate  millions of triggers per night which is expected to revolutionize our understanding of the astrophysical transient phenomenon.

In the time-domain astronomy era, the phenomena 
of short-lived astronomical transients on the timescale from seconds to hours are mysterious 
to be revealed and studied systematically.  
Bright or energetic events with such short timescales are  thought to result 
from the violent explosions in small radius, usually associated with
stellar outbursts, black holes, neutron stars or white dwarfs.  
Among these short-lived outbursts, one of the most interesting phenomena in the past 50 years is the gamma-ray burst (GRB), which manifests as a sudden outburst in gamma-ray band or hard X-ray due to either the death of massive stars or the merger of two compact objects (see the reviews by \cite{2004RvMP...76.1143P},\cite{2007ChJAA...7....1Z},\cite{2015PhR...561....1K} ). 
The latter  is confirmed to be one of the main electromagnetic counterparts of the gravitational wave events detected by the advanced LIGO/VIRGO detectors. Simultaneous observations of GRBs in multi-wavelength from the optical to the GeV bands enable us  
to obtain a full-band spectrum of GRB's prompt emission with a typical timescale of seconds, which sheds light on the physical mechanism of the relativistic outflows occurring immediately after the explosions. However, the studies in this field
have been greatly impeded due to the rare detection of GRB's optical prompt emission
in past decades. 
Although optical emission has been detected before the end of the prompt emission in a dozen GRBs (\cite{2019A&A...628A..59O}), the prompt phase has been entirely covered by such 
detection in only a few cases (e.g., \cite{2006Natur.442..172V,2014Sci...343...38V,2008Natur.455..183R,2023NatAs...7..724X}). 
Due to the very short duration of the prompt phase, a
dedicated facility with a cadence of seconds and very large spatial coverage is essential for catching and studying these particular    
optical transients. 

The Ground-based Wide Angle Camera array (GWAC), one of the main ground-based facilities of the Chinese-French \textit{SVOM} mission (\cite{Wei_2016arXiv161006892W}), is dedicated to detecting 
GRB's optical prompt emission down to a brightness of 15 magnitudes in 
the visual band.  

There are three main stages in the development of GWAC system.  In 2015, the prototype of GWAC,  Mini-GWAC, began operations. The Mini-GWAC is composed of 12 Apogee U9000X cameras
installed on six mounts. Within an operation of three years, a custom-designed real-time pipeline was developed to search for optical transients from images covering 4800 square degrees at a cadence of 15 seconds. This demonstrates the feasibility of searching for short-duration transients from high-cadence images over a very large field of view, such as the optical counterpart of gravitational waves (GWs, \cite{2020RAA....20...13T}).
The second generation of GWAC was installed in 2017 by increasing the camera's diameter and replacing the detectors by the e2v 4K$\times$4K CCDs. At this stage, a new pipeline system developed entirely in C language has been successfully implemented to meet real-time data processing requirements.  An automatic follow-up system was also established (\cite{Xu_2020arXiv200300209X}).
The third generation began in 2021. 
The detectors were replaced gradually with 4K$\times$4K CMOS sensors to enhance the survey cadence
thanks to the fast readout capability of CMOS technology.

Currently, GWAC could cover a joint sky coverage of about 3600 square degrees at a cadence of 3 seconds.
In addition to GRBs, such a high cadence survey allows us to potentially detect other
fast transients,  
such as the plausible bright optical emission associated with fast radio bursts and the very powerful superflares from ultracool stars, and the breakout of supernova.
This paper presents an overview of current status of the GWAC system, 
including the GWAC and the affiliated facilities.

The paper is organized as follows. 
The hardware and system structure of the GWAC system are described in Section 2.
Section 3 and 4 present the designed observation strategy, and Synchronized observations with \textit{SVOM}, respectively. 
The performance of astronomical and photometric calibration is reported in Section 5.
Sections 6 and 7 show the real-time data process and early scientific results obtained in 
the past 5 years. Finally, a summary and anticipated science in the forthcoming \textit{SVOM} era 
are discussed in Section 8.

\section{GWAC system}

The GWAC system, which is deployed at Xinglong observatory of National Astronomical Observatories, Chinese Academy of Sciences (NAOC),
consists of ten GWAC units and two follow-up telescopes, named F60A and F60B. The F60A and F60B telescopes are twin telescopes operated jointly by NAOC and Guangxi University. They are installed beside the array. Additionally there are  two more telescopes, F30A and F50A for the follow-ups. Both were jointly operated by NAOC and Huaibei Normal University  since November 2015 and January 2023, respectively. From September 2025, the two telescopes have ceased science operations at Xinglong Observatory. 

\subsection{GWAC subsystem}

GWAC system consists of 12 subsystems, specifically including:

\begin{enumerate}
    \item[1.] Dome Control System
    \item[2.] Mount Control System
    \item[3.] Focusing System
    \item[4.] Detector Control and Imaging System
    \item[5.] Real-time Data Processing System
    \item[6.] Fine Data Processing System
    \item[7.] Transient Source Identification System
    \item[8.] Follow-up Observation System
    \item[9.] Scientific Database
    \item[10.] Operational Status Monitoring
    \item[11.] Observation and scheduling
    \item[12.] Alert Reception and Dispatch
\end{enumerate}

\begin{table}[htbp]
\centering
\caption{GWAC System Parameters}
\begin{tabular}{|l|c|}
\hline
\textbf{Parameter} & \textbf{Specification} \\
\hline
Number of Dome & 2 \\
\hline
Number of Mount & 10 \\
\hline
Number of Camera & 40  (JFoV) \\
 & 10  (FFoV) \\
 \hline
Camera diameter & 18 cm (JFoV) \\
 & 3.5 cm (FFoV) \\
\hline
Pixel number of CMOS detector & 4096 $\times$ 4096 (JFoV) \\
 & 6248 $\times$ 4176 (FFoV) \\
\hline
Pixel scale & 8.37 arcseconds (JFoV) \\
 & 37.3 arcseconds (FFoV) \\
\hline
Total Field of View & 3600 sq. degrees (JFoV) \\
\hline
Exposure Cadence & 3 seconds \\
\hline
JFoV Limiting Magnitude & V16 mag @10 seconds \\
 & (under new moon) \\
\hline
Dedicated Follow-up Telescopes & 2 $\times$ 60 cm \\
\hline
Locations & Xinglong observatory \\
\hline
\end{tabular}
\label{GWAC_parameters}
\end{table}

\subsection{Dome}

There are two flat domes for GWAC,located in the north and south, respectively. All of them are installed at Xinglong observatory. 
Four GWAC units and the two follow-up telescopes are set in the north dome, while the remaining mounts are located in the south dome. 
The domes could be controlled independently. During the observations， the roof of each dome is fully opened, with one half sliding to the east and the other half to the west. This kind of flat dome is good for the survey with very large FoV instruments. 

\subsection{The Array of GWAC Units}

\begin{figure}
   \centering
     \includegraphics[width=5cm, angle=0]{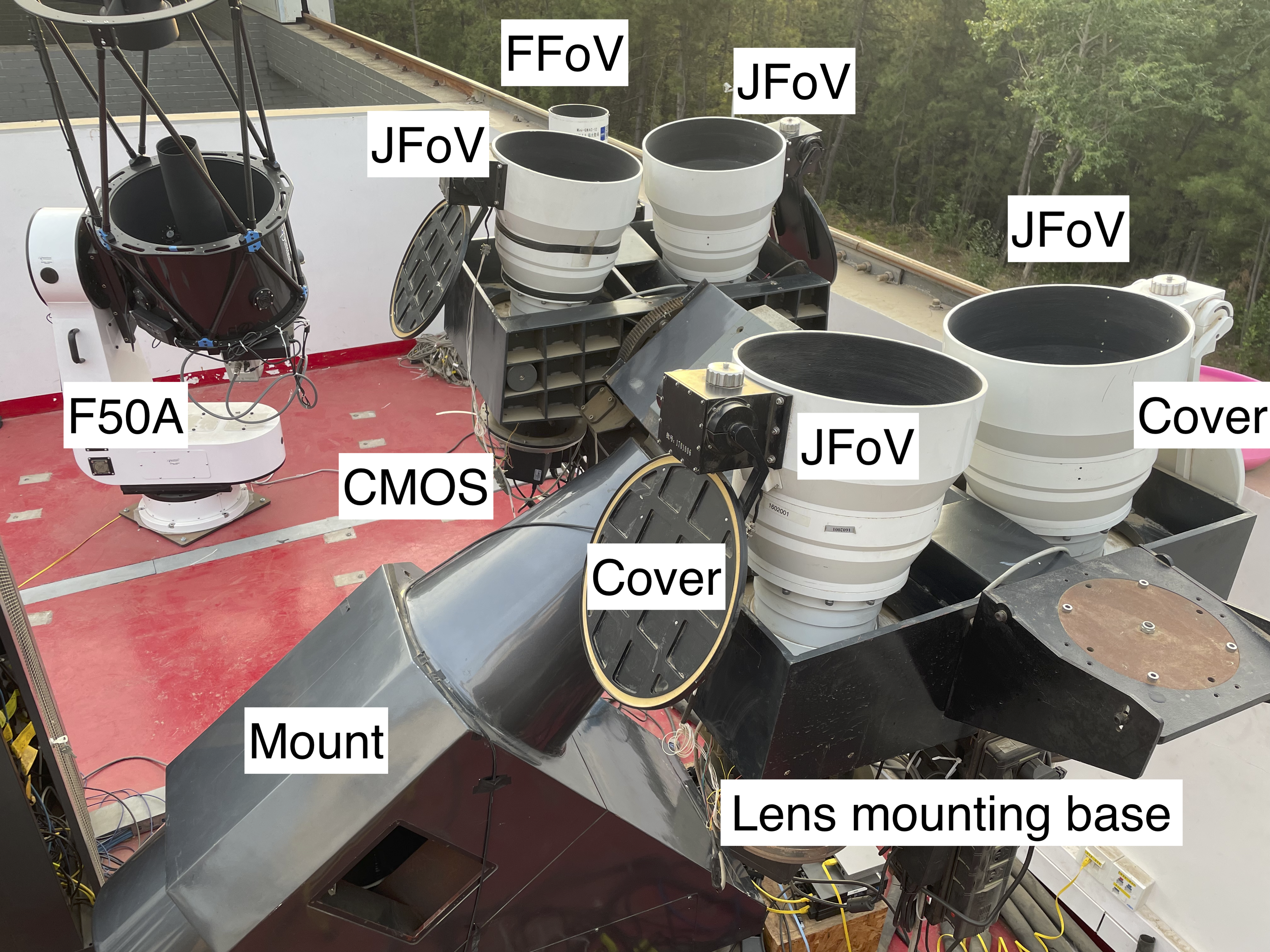}
   \caption{An overview of one GWAC unit. Four JFoVs and one FFoV are installed on one mount.
   Each JFoV, equipped with a 9 $\mu$m 4K$\times$4K CMOS, has a cover that can be removed and 
   recovered automatically by a stepper motor driver. }
   \label{GWAC_hd}
\end{figure}

The structure of each GWAC unit, consisting of an equatorial mount and a set of mounted cameras, 
is shown in Figure.\ref{GWAC_hd} and described briefly in this section. 
Table.\ref{GWAC_parameters} summarizes the main parameters of the GWAC system.

We refer the reader to \cite{2021PASP..133f5001H} for more details. 
The four JFoVs are installed on a lens-mounting module with a two-dimensional
V-shape benchmark whose deflection angle is fixed to be the field-of-view (FoV) of
individual JFoV. After an installation on the V-shape benchmark, the boresight of each 
JFoV is slightly adjusted to ensure that (1) there is no gap in sky coverage synthesized by 
any two adjacent JFoVs; and (2) the overlap is as small as possible to avoid significant 
loss of sky coverage.    
Each JFoV is a transmission telescope\footnote{http://www.forecam.com/} with a diameter of 18cm and a f-ratio of 1.2.
The optical design of the JFoV employs a total of eight lenses to provide a large FoV with negligible vignette.
An equipped 9$\mu$m 4K$\times$4K CMOS\footnote{https://www.tucsen.com/dhyana-4040-fsi-product/} 
detector results in  a pixel scale of 8.37 arcseconds and a FoV of 9.5$\times$9.5 square degrees.

Additionally, there is  a commercial SIGMA 50mm f-1.4 camera (hereafter FFoV ) mounted beside
the four JFoVs on each mount. The FoV of FFoV is large enough to cover the combined FoV by 
the corresponding four JFoVs. 
The detectors used for the FFoV is the ZWO ASL6200 CMOS\footnote{https://www.zwoastro.com/product/zwo-asi2600mc-air/}, which has 6248$\times$4176 pixels. A pixel size is 3.76 $\mu$m, resulting in 
a pixel scale of 37.3 arcseconds. 

During the survey each night, all the five cameras in each mount work simultaneously with a
cadence of 3 seconds. 
The FFoV plays two roles in the survey. One is to monitor the bright or brightened sources 
that are saturated in the JFoV images. The other is to provide a fast assessment of 
the pointing of the mount, with the well calibrated bias between the boresight of FFoV and the 
body-fixed coordinate system of the mount in advance.

\subsection{Automatic observation management system}
In addition to self-searching for all short-lived astronomical events by monitoring millions of stars, GWAC can perform a Target of Opportunity (ToO) follow-up of gamma-ray bursts from the \textit{SVOM} (\cite{Wei_2016arXiv161006892W}), The Neil Gehrels Swift Observatory 
 (\cite{2004ApJ...611.1005G}), the Einstein Probe
(\cite{2022hxga.book...86Y}),  or the Fermi missions, or gravitational waves from the LIGO/VIRGO detectors, thanks its large FoV. To meet these scientific requirements, we have designed several observing modes and the strategies, including a joint strategy with several GWAC telescopes and their dedicated telescopes, especially for ToO observations of gravitational wave (GW) events with a very large skymap characteristic for their poor localisation. To carry out these observations, a custom-designed system named the Automatic Observation Management (AOM) was developed with capabilities including target management, dynamic scheduling, automatic network-wide broadcasting, and image processing. The AOM system connects the individual telescopes in the network and organises all the associated operations with good performance. 
With its modular design, the AOM can manage more telescopes as new units are installed. More detailed information can be found in the literature \cite{2021PASP..133f5001H}.

\subsection{Automatic focus}
When observing with a ground-based optical telescope, the final image quality, especially for the  encircled energy, is affected by many factors, including seeing conditions, ambient temperature, lunar phase and angular distance, light pollution and clouds. Due to the large pixel scale, 
the image quality of GWAC is not sensitive to seeing as the telescopes with small pixel scale, but it is more sensitive to ambient temperature, which typically varies by several degrees gradually over a long-term trend each night.  

Since the 50 cameras of GWAC working simultaneously, comprising 10 FFoV and 40 JFoV, it is not realistic to make the refocus them manually one by one each night. In order to maintain the image quality close to its best level, which is essential for the transient detection capability, 
each of the cameras is equipped with a custom-developed automatic focusing system (Huang et al., 2015).
The focusing system works with a stepper motor (without a position detection sensor) 
that is controlled in a closed loop with the image PSF profile, as assessed by full width at half 
maximum (FWHM) in real time during the observations.

\begin{figure}
   \centering
  \includegraphics[width=6cm, angle=0]
  {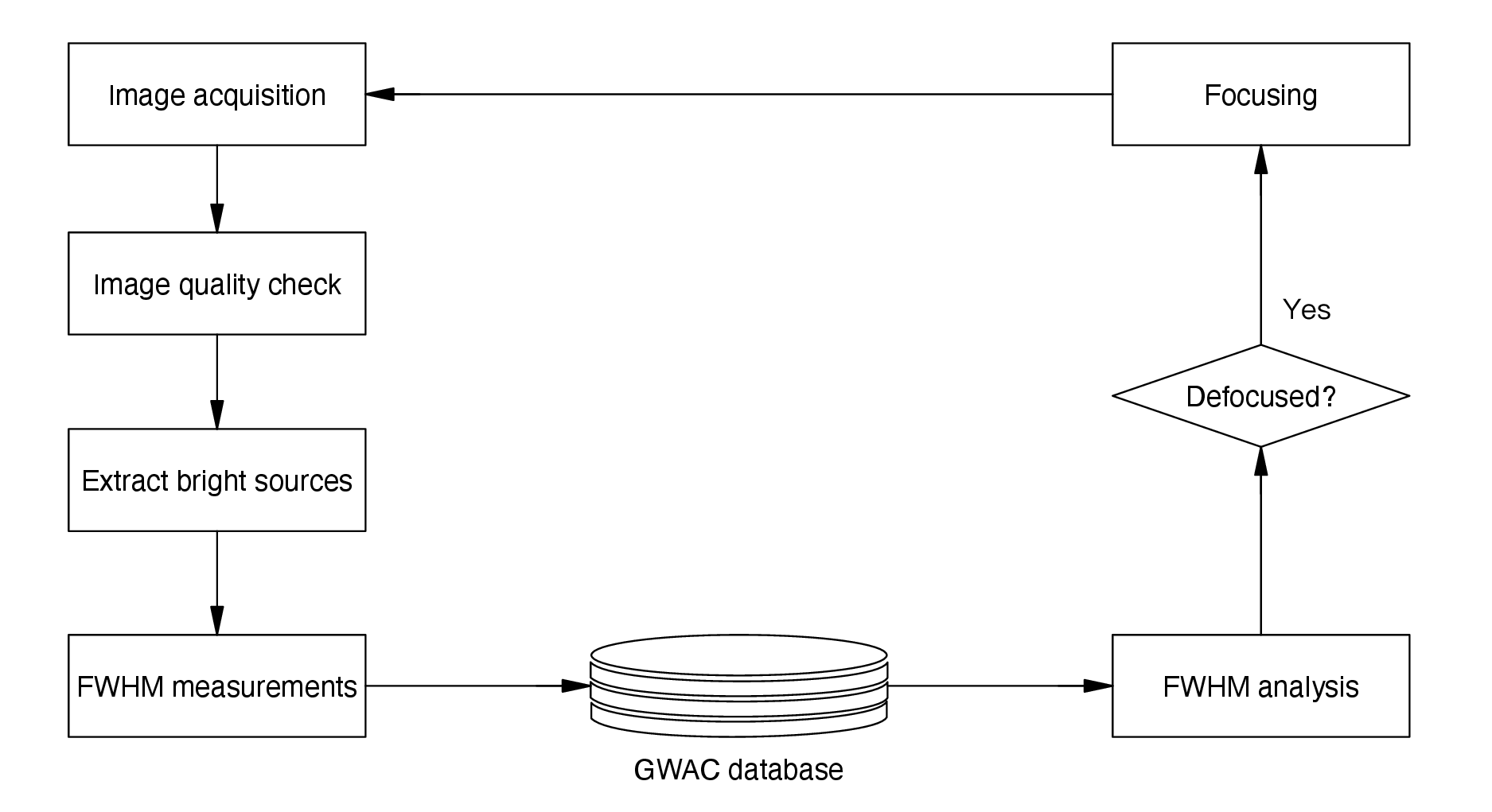}
   \caption{The loop of the automatic focusing system of GWAC. 
   } 
   \label{GWAC_focusing}
   \end{figure}

\begin{figure}
   \centering
  \includegraphics[width=7cm, angle=0]
  {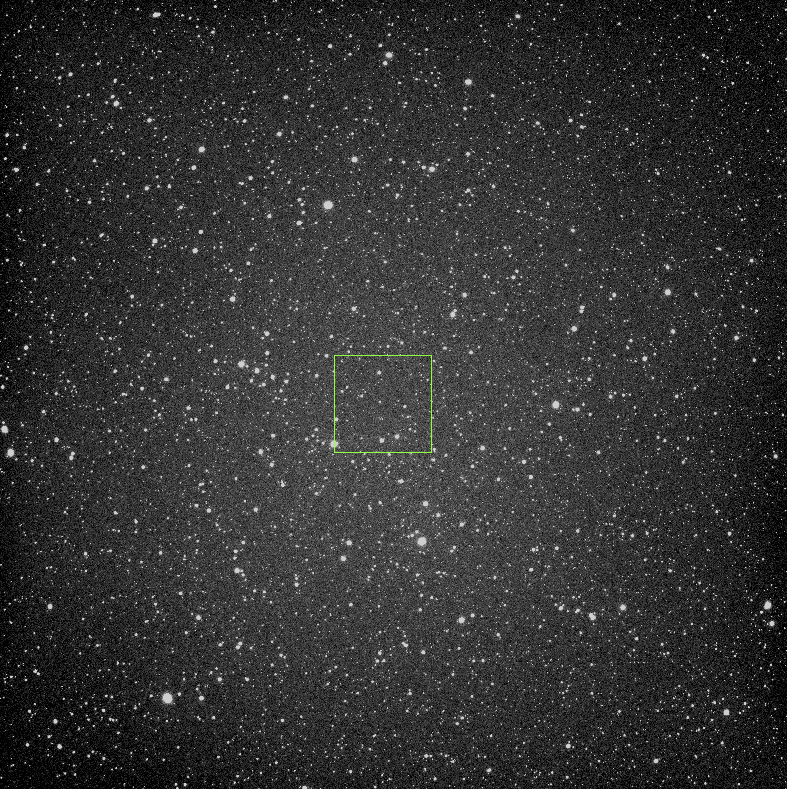}
    \includegraphics[width=7cm, angle=0]
    {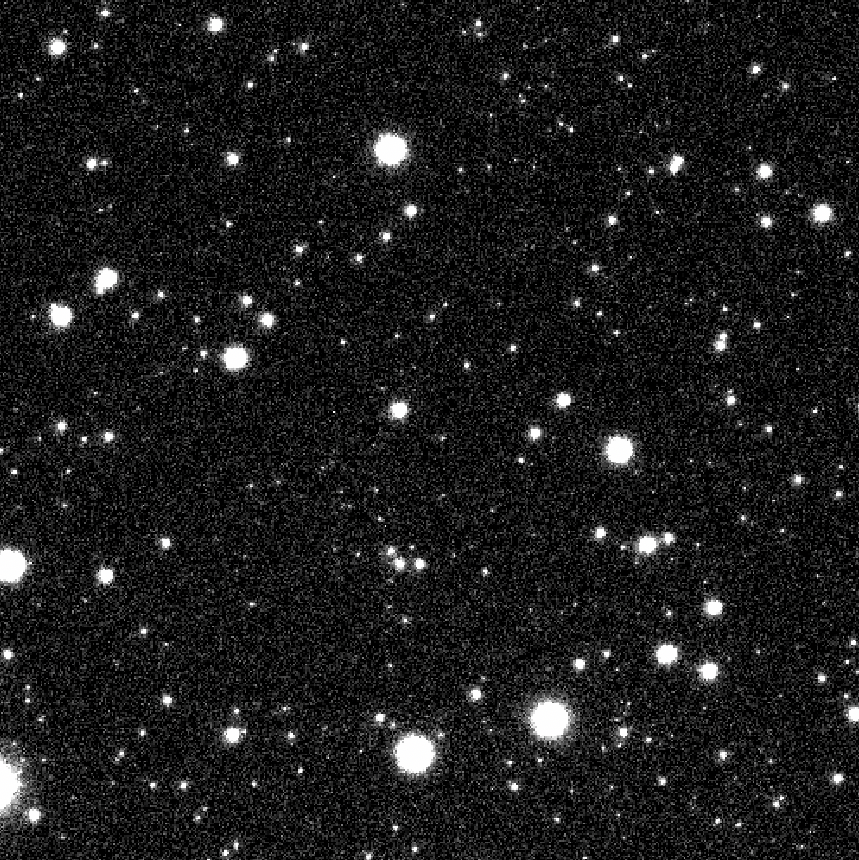}
   \caption{A raw GWAC image, obtained with a CMOS detector on 2023 February 15, is shown without any corrections.  \it Upper panel: \rm The full-frame image with a total pixel size of 4096$\times$4096, yielding a FoV of 9.5$\times$9.5 square degrees. \it Bottom panel: \rm A zoomed image of the green box centered at the upper panel with a size of 500$\times$ 500 pixels.} 
   \label{GWAC_image}
   \end{figure}

The loop of the system is schematically shown in Figure.\ref{GWAC_focusing}. 
After extracting bright sources in a frame, the FWHMs of these sources are measured 
by using IRAF\footnote{IRAF is distributed by NOAO, which is operated by AURA,
Inc., under cooperative agreement with the U.S. National Science
Foundation (NSF).}, which leads to a mean value, and an associated uncertainty for each frame.
A dedicated pipeline has been developed to monitor the variation of the mean values, and to 
initiate or halt focus adjustment when the mean values deviates from or remain within a given threshold relative to the optimal value. Several iterations are typically required to finish 
focus procedure by achieving a focusing accuracy of 0.05 pixels. 
An image obtained from the automatic focusing system is shown in Figure.\ref{GWAC_image}, where the FWHM is measured to be 1.8$\pm$0.2 pixels across the entire field. 
More detailed information based on the automatic focus system can be refereed (Huang et al., 2015). The new system with the encircled energy ratio (EER) has also been developed recently, where the EER is a ratio of the energy within a circular aperture of fixed radius to the total energy from the full aperture. The EER is similar with the FWHM which serves as an indicator of the image quality but is more robust during the observations.

\subsection{Dedicated follow-up system}

GWAC typically generates millions of internal triggers  each night in real time. After a series of filters, only a few dozen candidates needed to be quickly validated and identified by duty 
astronomers in order to initiate and coordinate rapid photometry in multi-wavelength,and spectroscopic observations. 
A real-time automatic transient validation system  
has been developed to reduce the delay
between a GWAC trigger and subsequent follow-up observations (\cite{Xu_2020arXiv200300209X}) 
using two F60A and F60B telescopes. The two telescopes are identical, each equipped with an Andor 2K$\times$2K CCD, providing a field of view (FoV) of 19.1 arcminutes.  
Observations over the past four years demonstrate that this system 
works well, including automatic validation of the candidates, adaptive light-curve sampling for in multiple bands, and rapid distribution of follow-up results to the  duty astronomer via a mobile client. Duty astronomers need only pay attention to those candidates automatically validated by the system.

\section{Observation strategy}

There are two types of observations, i.e., routine survey and ToO observations. 
The entire available sky is divided into hundreds of fields  in advance,
with a size of each field determined by the total FoV of one unit, i.e., about 356 square degrees.
During the survey, all the units point to different sky fields with dynamically allocated  
priorities that are evaluated by the AOM system by taking into account the elevation, 
Moon phase, angular distance from the Moon, and the Galactic plane (\cite{2021PASP..133f5001H}, \cite{Han+etal+2026}). Once an external trigger with a high priority is received, e.g. \textit{Swift}/BAT or \textit{Fermi}/GBM triggers,
GWAC switches to ToO observation in response to these alerts by following the pointing strategy used in the survey. 
The primary goal is to detect the  optical emission before, during, and after the period of high-energy prompt emission. Thanks to its large FoV of 3600 square degrees, if GWAC already covers the error box of an alert, it does not need to repoint. If not, GWAC can slew to the alert at a rotate rate of 2 degrees per second. Meanwhile, if the localization uncertainty is better than 10 arcminutes, F60A and F60B, which have better detection capabilities, will conduct follow-up observations, and GWAC will not observe the target. 
For both observation modes, in order to search for transients in real time, 
a reference image and the corresponding reference catalog in each field 
must be available in advance. 
Because of GWAC's large pixel scale, these reference image and catalogs 
are extracted from GWAC own observations. 

\section{Synchronized observations with SVOM}

\begin{figure}
   \centering
  \includegraphics[width=8cm, angle=0]
  {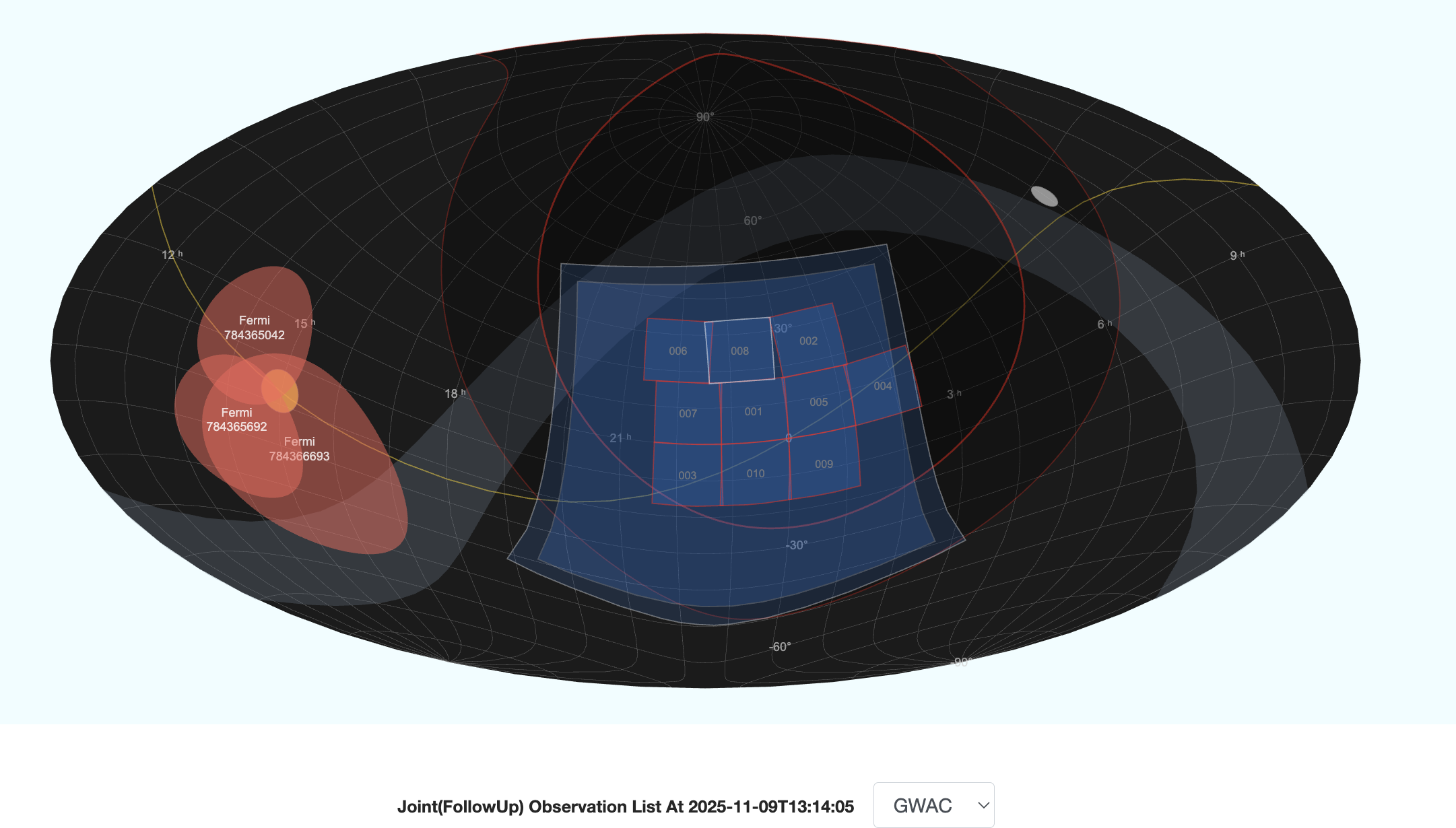}
   \caption{The joint observations at 2025 November 09, by GWAC and the \textit{SVOM}, ten mounts are scheduled to the sky within the FoV of Eclairs. Here the alerts from Fermi at the same time and the Galactic plane are also displayed.} 
   \label{svom-gwac}
   \end{figure}

\textit{SVOM} satellite in orbit follows an anti-solar strategy by observing the sky with a "B1" law during its survey.  The B1 law is defined primarily by  taking into account the Sun constraints, Galactic plane, and other factors, in order to maximize the GRB detection rate, as well as  follow-up observations with telescopes on-ground. As a result, \textit{SVOM} points to the southern part of the sky during the summer and to the northern part during the winter. At Xinglong observatory in China, the weather is \textbf{more favorable} for the observations in  winter.
In order to increase overlap between the sky observed by \textit{SVOM} and GWAC, the schedule of the GWAC each night is based on the  observation plans generated for \textit{SVOM}. The sky fields in the \textit{SVOM} FoV,  also are simultaneously visible for GWAC would have higher priority during the scheduling. Figure.\ref{svom-gwac} shows the joint observations, as an example, during the observations at 2025 November 9. Ten mounts of GWAC covering about 3600 square degrees are scheduled to cover the sky within the FoV of \textit{SVOM}/Eclairs.

\section{Calibrations}
The astrometric and photometric calibrations of the GWAC images are described in this section.

\subsection{Astrometric calibration}

Astrometric calibration aims to determine a 
transformation between the detector coordinates (X, Y) and the celestial coordinates (R.A. and Decl.). A custom-designed pipeline was developed for the GWAC images based on the \it IRAF.ccmap \rm package, which matches the brightest stars extracted from each image with 
the reference catalogs, i.e., the Tycho-2 and USNO B1.0 catalogues. 
The astrometric solution is then parameterized by a 
2-dimensional Legendre function via a $\chi^2$ minimization.  The order of the function is configurable and depends on the number of the extracted bright stars.

\begin{figure}
   \centering
  \includegraphics[width=6cm, angle=0]
  {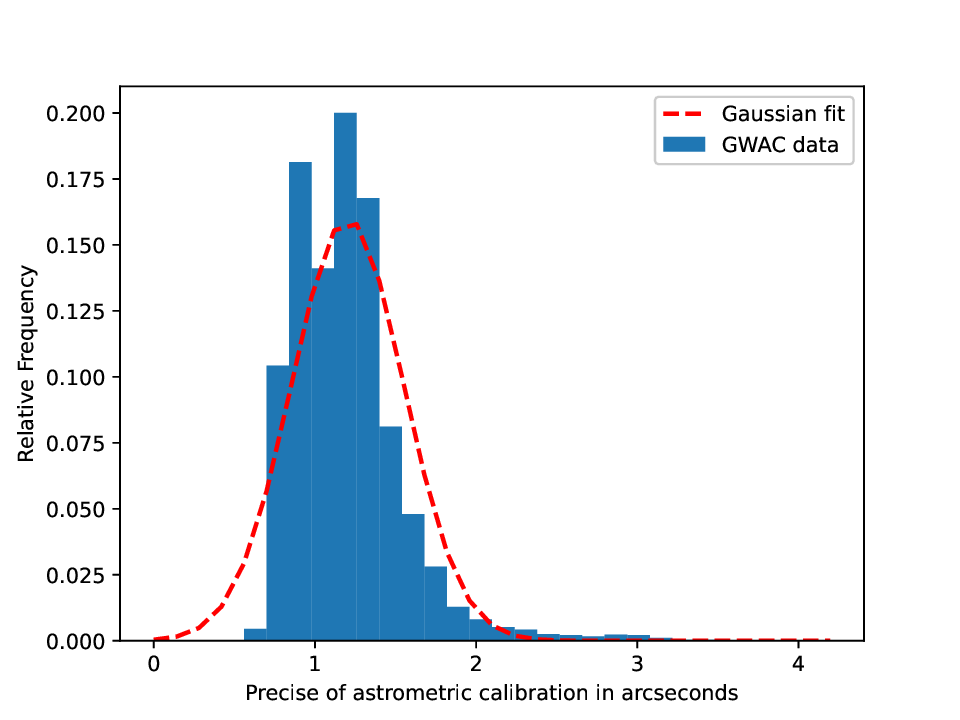}
   \caption{Accuracy distribution  of astrometric solution for 33,7740 GWAC images taken from November 11th 2020 to 2023 June 17th. The best-fit Gaussian function to the distribution is over plotted by the red line, which returns a mean value of $1.2\pm0.4$ arcseconds.} 
   \label{long-term-astrometric}
   \end{figure}

Figure.\ref{long-term-astrometric} shows the accuracy distribution of the determined
astrometric solution for 33,7740 GWAC images obtained from 2020 November 11th to 2023 June 17th.
The accuracy of astrometric solution is defined as 
$\Delta = \sqrt{\Delta_x^2+\Delta_y^2}$, where $\Delta_x$ and 
$\Delta_y$ are the standard deviation of residuals in CCD's X and Y directions, respectively.
A Gaussian fit to the distribution gives a mean value of $\Delta = 1.2\pm0.4$ arcseconds, 
suggesting a good performance of our astrometric calibration. 

\begin{figure}
   \centering
  \includegraphics[width=7cm, angle=0]
    {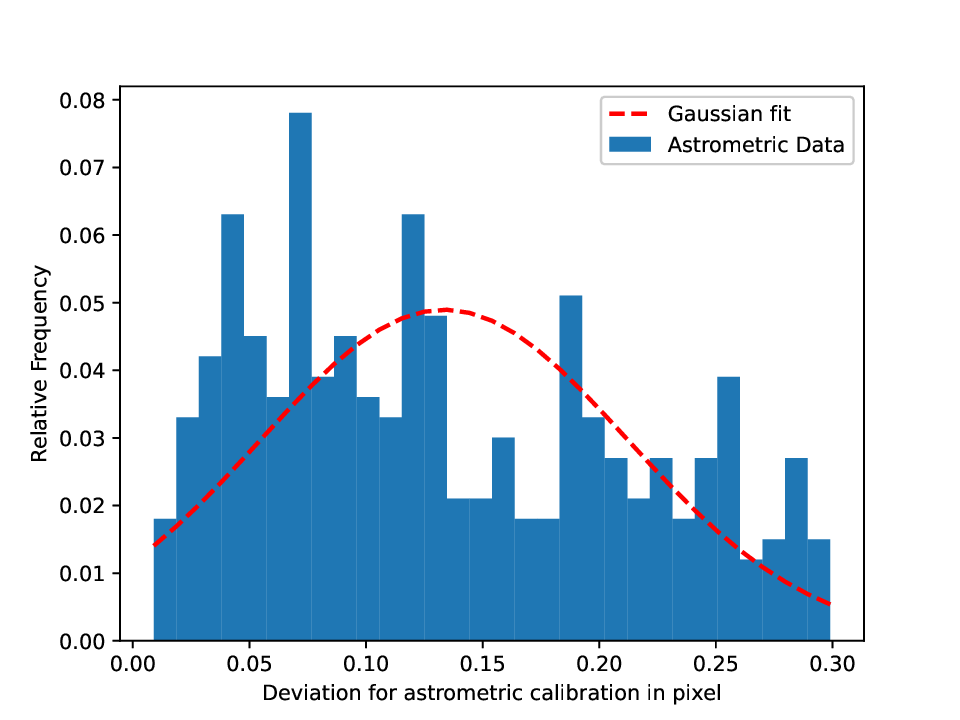}
    \includegraphics[width=7cm, angle=0]
      {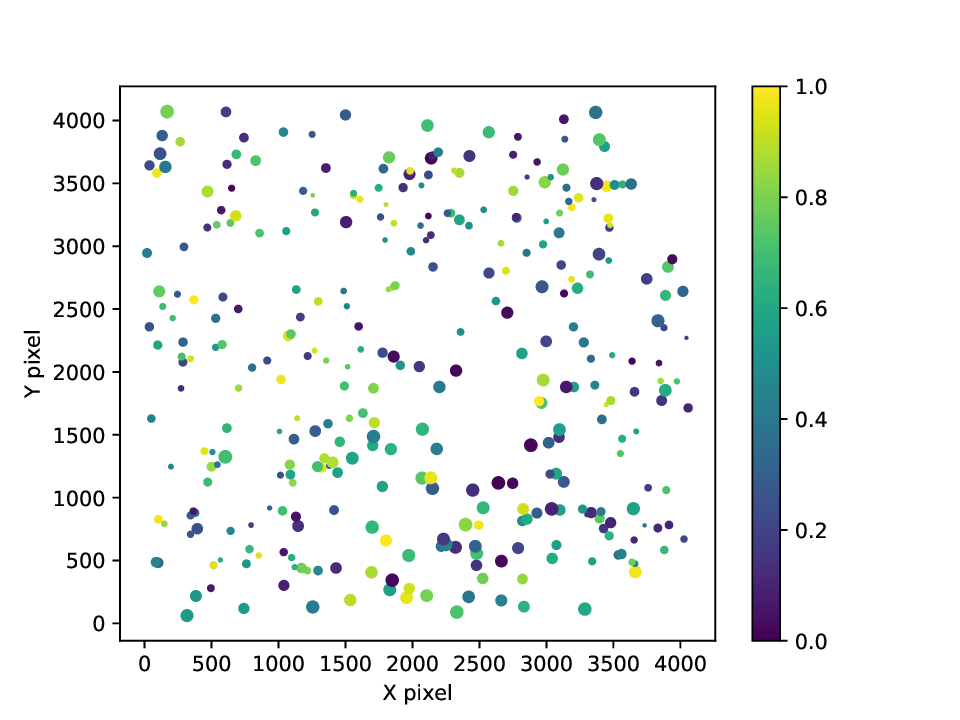}
   \caption{The performance of the astrometric calibration assessed by the bright stars with 
   $\mathrm{S/N}>20$ in a GWAC image. 
   \it Upper panel: \rm a distribution of the difference between the measured and
   predicted positions on CCD (see the main text for the details), which yields a mean 
   and maximum difference of $0.13\pm0.08$ and 0.3 pixels, respectively. \it Bottom panel: \rm 
   The differences in pixels, shown by the different color level, as a function of positions on CCD.} 
   \label{astrometric}
   \end{figure}

As an additional test, Figure.\ref{astrometric} compares the measured and predicted positions 
on the detector of all the bright stars with signal-to-noise ratio $\mathrm{S/N}>20$ in a 
typical GWAC image.  
The measured positions are obtained by \it SExtractor \rm (\cite{1996A&AS..117..393B}).
The errors of these measurements are negligible since the used stars are bright enough.     
Based on the astrometric solution determined as described above, 
the corresponding predicted positions are calculated from the celestial coordinates listed in 
the USNO B1.0 catalogue. The left panel of the figure presents a distribution of the 
difference between the measured and predicted positions, which yields a mean value of 
$0.13\pm0.08$ pixels and a maximum deviation of 0.3 pixels. 
This mean value is slightly larger than the value of 0.08 pixels when 
the Tycho-2 catalogue is used. This difference is partly due to the different precision of the coordinates of the two catalogues, or to the correction of the proper motion of nearby stars. The right panel of the figure shows the position difference as a function of position on detector. One can find that there is not any obvious clustering effect for the
position difference, which suggests the astrometric solution has a nearly uniform accuracy level across the entire frame.

\subsection{Photometric calibration}

GWAC does not work with a standard photometric system, because of its wide wavelength coverage.
The front lens of GWAC is coated to allow photons in the 500-800 nm range to pass through to 
reduce chromatic aberration. The GWAC photometric results are compared with those from the standard 
photometry systems using the following procedure.

\begin{figure}
   \centering
     \includegraphics[width=7cm, angle=0]
     {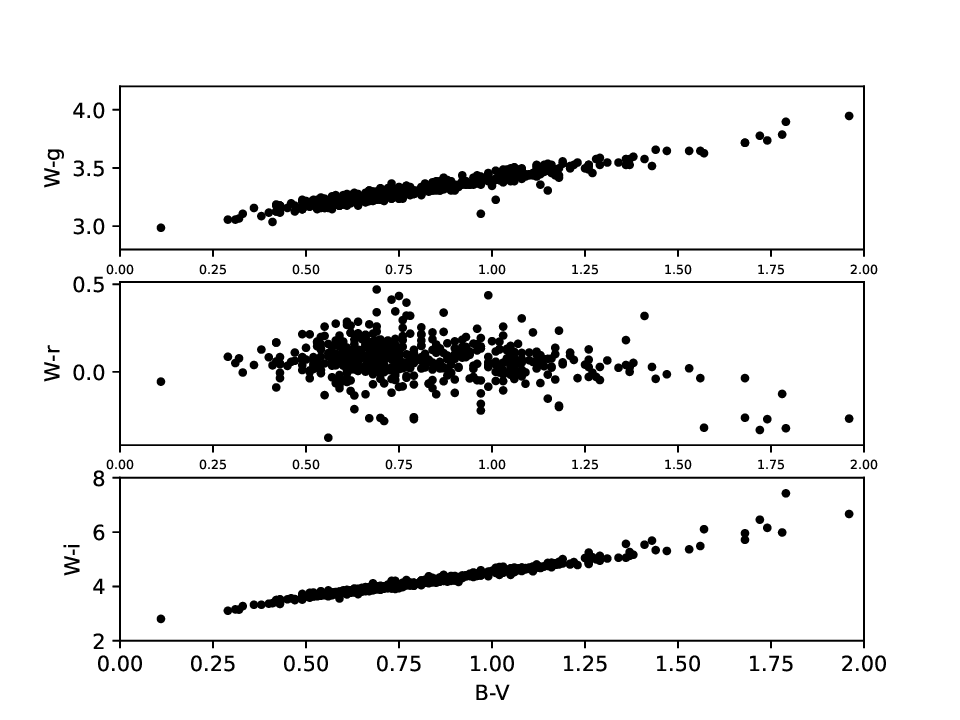}
  \includegraphics[width=7cm, angle=0]
  {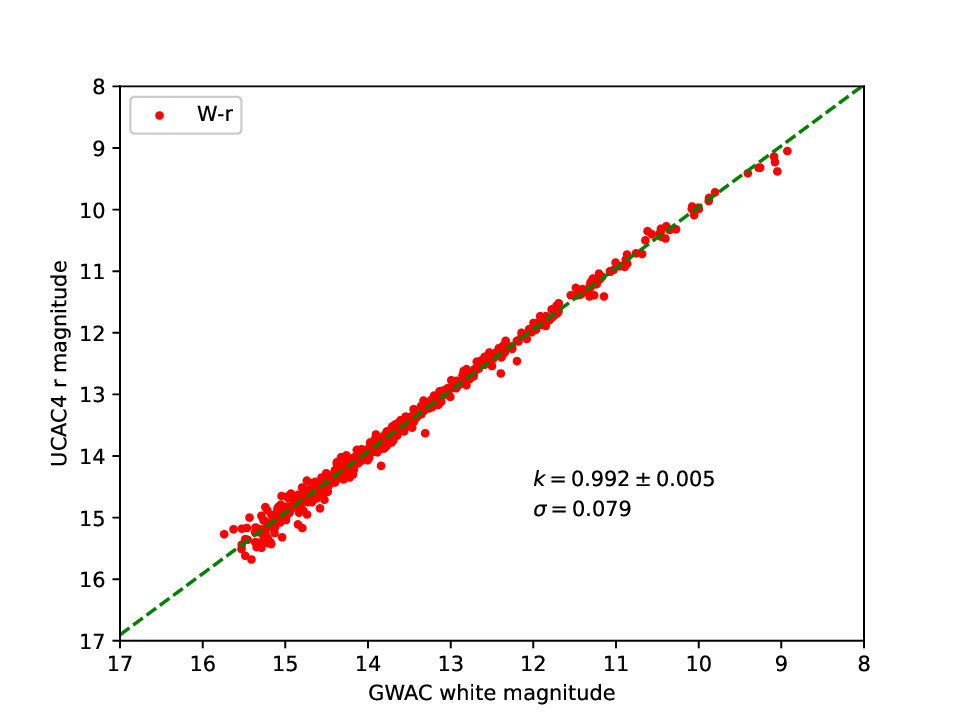}
   \caption{A comparison between GWAC white-band magnitude ($W$) and UCAC4 multi-wavelength magnitudes. \it Left panel: \rm The colour-colour diagrams suggesting that the UCAC4 $r$-band magnitude is more comparable to GWAC white-band magnitude than others. \it Right panel: 
   \rm GWAC white-band magnitude plotted against UCAC4 $r$-band magnitude. The best linear fit 
   is over plotted by the green line, which returns a coefficient factor of 
   $k=0.992\pm0.005$ and an uncertainty of $1\sigma=0.079$.  }
   \label{GWAC_W-r}
   \end{figure}

First, the astrometric solution determined in Section 4.1 is applied to each frame,
after bias, dark and flat-field correction. The GWAC photometric results are then 
compared with the UCAC4 catalog (\cite{2013AJ....145...44Z}) that is adopted for two reasons.
(1) The catalogue contains over 113 million objects and is nearly complete down to 
a brightness of 16 magnitudes in the visual band, which is close to the detection limit of 
GWAC; and (2) it provides magnitudes in five bands ($B, V, g, r, i$) for each object.  
Based on the images obtained on February 5th, 2022,
Figure.\ref{GWAC_W-r} compares the GWAC white-band magnitude ($W$) with the UCAC4 multi-band magnitudes, showing that the GWAC photometric system is very similar to the 
UCAC4 $r$-band photometry.
The colour-colour diagrams in the left panel, in fact, indicate that the $r$-band magnitude is 
more comparable to the GWAC white-band magnitude than the others, as the color $W-r$
is nearly independent on spectral types (i.e., $B-V$). $W$ is plotted as a function of
$r$ in the right panel. A linear fitting yields a best-fit relationship:
\begin{equation}
   r=(0.992\pm0.005)W+b_0
\end{equation}
with an uncertainty of 0.079 at the 1$\sigma$ level.  

Figure 9 shows how the photometric uncertainty increases with magnitude, where the uncertainty
is defined as the difference between the calibrated $r$-band magnitudes measured in 
two consecutive images. One can see from the figure that the variation of 
median of the photometric uncertainty (the solid blue line) generally follows that 
of the magnitude uncertainties ( red points) measured for each object by SExtractor in 
either of the two images, which indicates that the scatter of the photometric uncertainty ( 
gray points) is primarily driven by the S/N for each source across the brightness distribution. In addition, it can be seen that the photometric uncertainty is $\sim2\%$ at
at the bright end, which represents the systematic level.

Based on the photometric calibration, the magnitude limits are investigated 
for a single GWAC frame as well as stacked image obtained at 2022 February 05th. The images with an effective
exposure times of 10 sec, 100 sec and 540 sec are shown in Figure.\ref{GWAC_diff_exposure}. 
The corresponding 5$\sigma$ upper limits in the $r$-band are determined to be 15.3, 15.9 and 
16.5 magnitudes.

\begin{figure}
   \centering
  \includegraphics[width=5cm, angle=0]
  {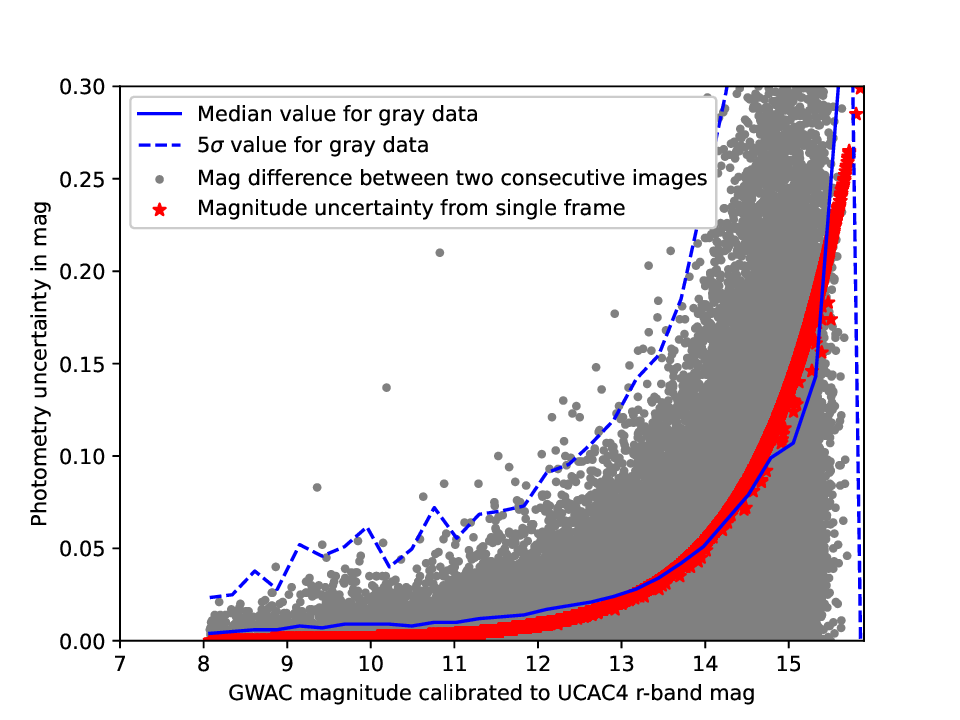}
   \caption{The photometry uncertainty plotted against magnitude. The gray data donates the magnitude difference of the same source in two consecutive frames. The solid blue line and dashed blue line are the median and the 5$\sigma$ confidence level, respectively. The red points are the values measured in single frame by the SExtractor.} 
   \label{GWAC_mag_error}
   \end{figure}

\begin{figure}
   \centering
  \includegraphics[width=2cm, angle=0]
  {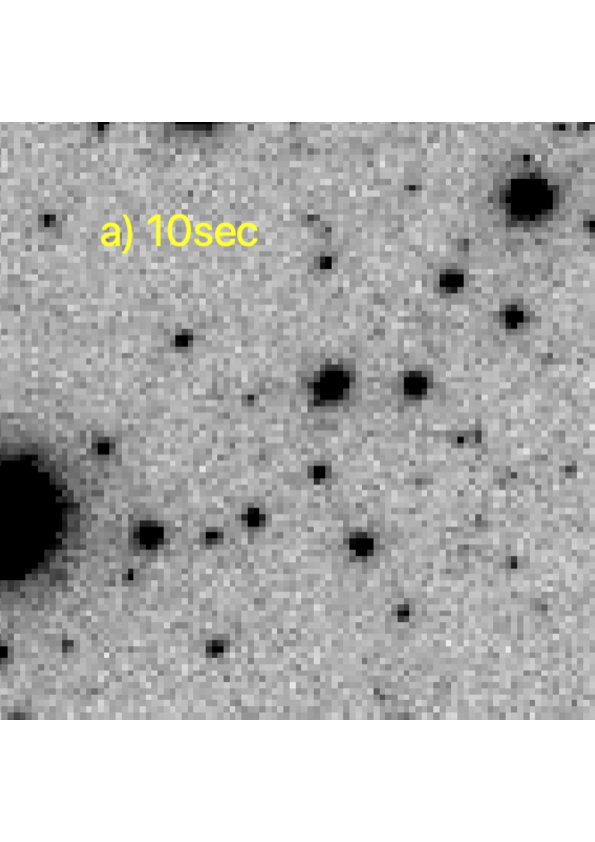}
  \includegraphics[width=2cm, angle=0]
    {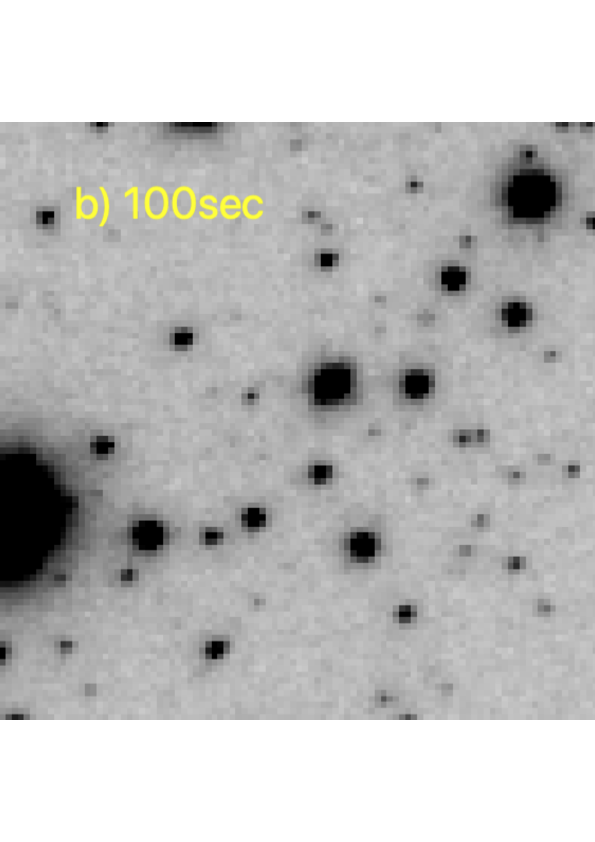}
  \includegraphics[width=2cm, angle=0]
    {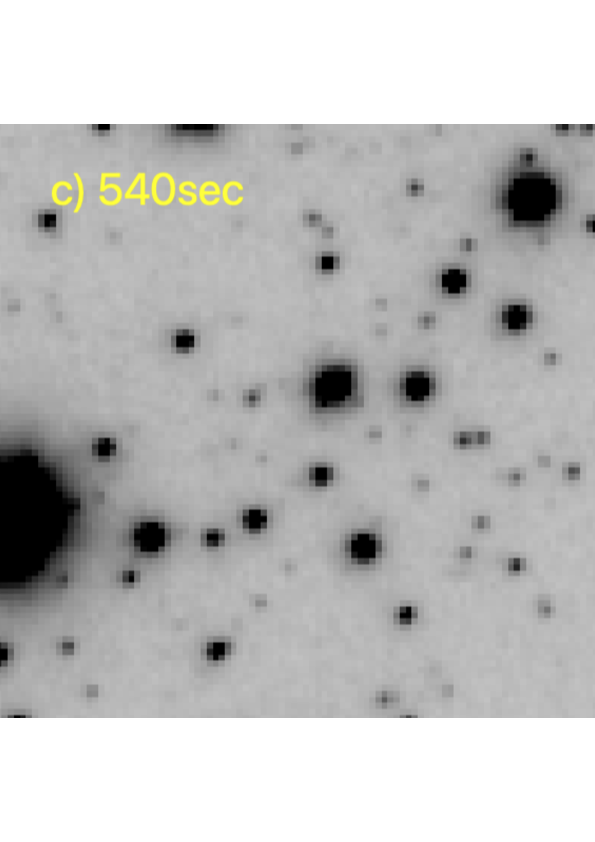}
   \caption{An illustration of subimages with different effective exposure times. 
   The panel a) is a single frame with an exposure time of 10 seconds, while panels b) and c) are generated by stacking 10 and 54 frames, respectively.} 
   \label{GWAC_diff_exposure}
   \end{figure}

\begin{figure}
   \centering
  \includegraphics[width=6cm, angle=0]
  {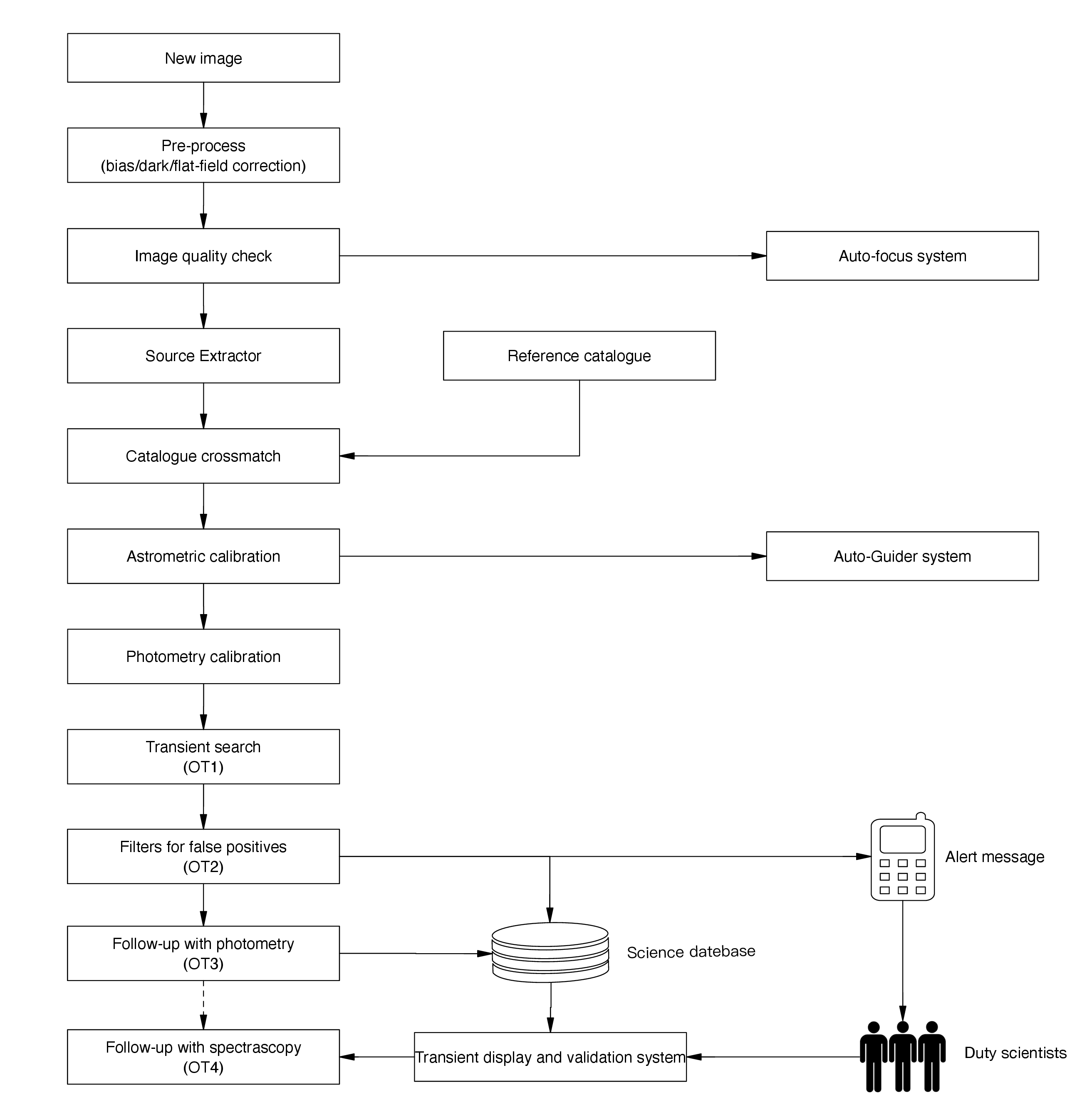}
   \caption{The brief summary of workflow for the real time pipeline via catalogue crossmatch method. The main functions during the process are presented with the aiming to search for transients from GWAC images, and to validate as well as further studies via rapid follow-ups with multi-wavelength photometry and spectroscopy.} 
   \label{pipeline workflow}
   \end{figure}

\section{Search for transient in real-time}

The main scientific motivation for GWAC is to capture the entire explosive process of short-lived astronomical events on  timescales of seconds to minutes, such as gamma-ray bursts and stellar flares.  
From the winter of 2023, all CCDs were replaced with CMOS sensors. The temporal resolution of the survey were also updated from 15 seconds (10 seconds plus 5 sec readout) to 3 seconds, thanks to the fast readout of CMOS with a low readout noise of about 3.4e. 

A \textbf{custom-developed} real-time catalogue cross-matching pipeline  was developed to search for the short-duration transients, taking into account the following considerations. First, the science what we currently focus on  primarily involves transients with duration from seconds to minutes. These transients need to be followed up quickly to explore the characteristics of the brightness evolution in more detail, which requires a real-time pipeline for processing all images. Second, the amount of data generated in total is very large, an order of one Terabyte B per night. it is very challenging to reprocess all this data off-line.
Third, it is also very challenging to keep all the raw data on our server for years. Only those data related to some \textbf{objects of interest} can be stored for a long time. Fourth, useful information could be derived from the data processing pipeline, including the absolute pointing precise, the accuracy of tracking and the image quality. All these derived from science images could be used to evaluate the performance of the system, and serve as inputs to the operations loop.

The pipeline workflow is shown in Figure.\ref{pipeline workflow}. When a new science image is acquired, the image quality check is performed first, followed by the bias, dark and flat field corrections. A source catalog is derived from the new science image using SExtractor (\cite{1996A&AS..117..393B}), which is then matched to the reference catalog, including the position and flux. The reference catalog is a combined catalog derived from USNO B1.0 and the reference image which is obtained in advance with the same GWAC detector under good observing conditions.

Once all the above matches are completed, any uncatalogued sources are searched in the science image with a threshold of S/N $>$3 by comparing with the reference catalogue. Only those  detected for at least twice in any five successive images are then labeled as OT1.  Several filters has been developed based on the characteristics of the instruments to filter out the false positives from the OT1 candidates. For GWAC images, these false positives include several main types. The first is the hot pixels or clusters that are not well corrected in dark images. The second are ghosts from very bright sources located in or very close to the FoV. The main characteristic of ghosts is that their PSF is typically wider than those of nearby stars. The third is false detections caused by the crosstalk between two or more objects whose positions are very close to each other. Additional false positive candidates result from  fluctuations in a very bright background especially when the pointing is too near a full Moon or when there are some clouds which make the background in the whole FoV not uniform. An inaccurate catalog match could also generate some "new" sources, for example a bad position match between catalogues across an entire image. Finally hundreds of  artificial objects or minor planets or comets may also be misidentified as OT1.

Once a candidate has passed all filters, it is labeled to be OT2. OT2 candidates are then sent to F60A/B for a quick follow-up to confirm their nature with two snapshots of 30 seconds exposure time each in  R-band. The typical delay time of the follow-ups from the first detection by GWAC is about 2 minutes (\cite{Xu_2020arXiv200300209X}). The typical detection capabilities for each image by F60A/B is about 17th magnitude, which allows us both to confirm the candidate, and to give a more precise localization of the source  (pixel scale is 0.56 arcseconds), as well as the brightness measurement. 

A real-time pipeline has also been developed for the images obtained by F60A/B, enabling observation results to be obtained within a few seconds. Once a candidate is confirmed, a more intensive but flexible multi-wavelength follow-up will be performed automatically to monitor the light curve evolution with F60A/B. 
During those follow-ups, both the exposure time and the cadence for each frame are adjusted dynamically by taking into accounts the brightness and the temporal decay of the light curve (\cite{Xu_2020arXiv200300209X}). A real-time alert is sent to the duty scientists  via the dedicated tools for human inspection or to further optimize the observing strategy if necessary. Meanwhile, if the source is bright enough (R$>$16 mag), a quick trigger can be sent by duty scientist to 2.16 m optical telescope at Xinglong Observatory for time-resolved spectroscopy with a predefined strategy.  

Typically, millions of OT1 candidates are detected each night, from which hundreds of OT2 candidates are generated. After rapid follow-up by F60A/B, only a few are confirmed each night as real astronomical events of interest. Most of these are stellar superflares (e.g., \cite{2024MNRAS.527.2232X,2023RAA....23a5016L}). 

In addition, a new pipeline based on the image difference is under development. The pipeline is designed to focus on the transients with a duration of minutes. During the procedure, the consecutive images are stacked with an effective exposure time of 30 seconds. The number of images to be stacked is configurable. The template image for the difference is selected with a gap of 10 minutes earlier. Due to the short time gap, the image quality and the background are very similar. This pipeline has been running in real time since 2025 October, with good performance on the detection of short duration transients, including stellar flares (Xu et al., in preparing). 

\section{Early sciences}
 
Many sciences have been achieved by the GWAC system over the last few years, even though it is during the procedure of the building.  
Due to its high cadence,
these studies focus on the phenomena of fast optical transients, including 
gamma-ray bursts, fast radio bursts, superflares of late-type stars.
A brief summary of these results is given in this section.

\subsection{Gamma-ray bursts}
Gamma-ray bursts (GRBs) are sudden outbursts of gamma-ray emission and are thought to be the most violent explosions in the universe after the Big Bang. Since the first report (\cite{1973ApJ...182L..85K}) about 50 years ago, our understanding of GRB physics has been greatly advanced, especially in the last 20 years since the launch of the \it Swift \rm mission (\cite{2004ApJ...611.1005G}) which can quickly provide a well-localized GRB trigger for ground-based telescopes. A widely accepted global picture of the GRB has been established, showing that GRBs originate from either the collapse of massive stars or the merger of compact objects. Once a new black hole or a new massive magnetar is born, a highly relativistic jet is lunched and the short-lived prompt emission is produced inside the jet by an internal interaction of the jet, 
while a long-lived multi-wavelength emission is produced, which is usually called the afterglow,
by the interaction between the jet and circumburst medium.   

Although comprehensive studies of the afterglow in both temporal and spectral domain have been carried out over the past decades, multi-wavelength observations of the short-lived prompt emission
(e.g., \cite{2005Natur.435..178V,2006Natur.442..172V,2008Natur.455..183R}) remain rare for at least two facts. 
First, the location and time of a new GRB cannot be predicted. Second,
most ground telescopes work in a ``trigger-follow-up'' mode due to their small FoV.
Instead of the ``trigger-follow-up'' mode, simultaneous observations in multi-wavelength by facilities with high cadence and a 
very large FoV
are essential for exploring the short-lived prompt emission.
GWAC was built following this motivation, and the primary science goal is to detect the optical emission before the end of the prompt emission. 

\begin{figure}
   \centering
    \includegraphics[width=7cm, angle=0]
    {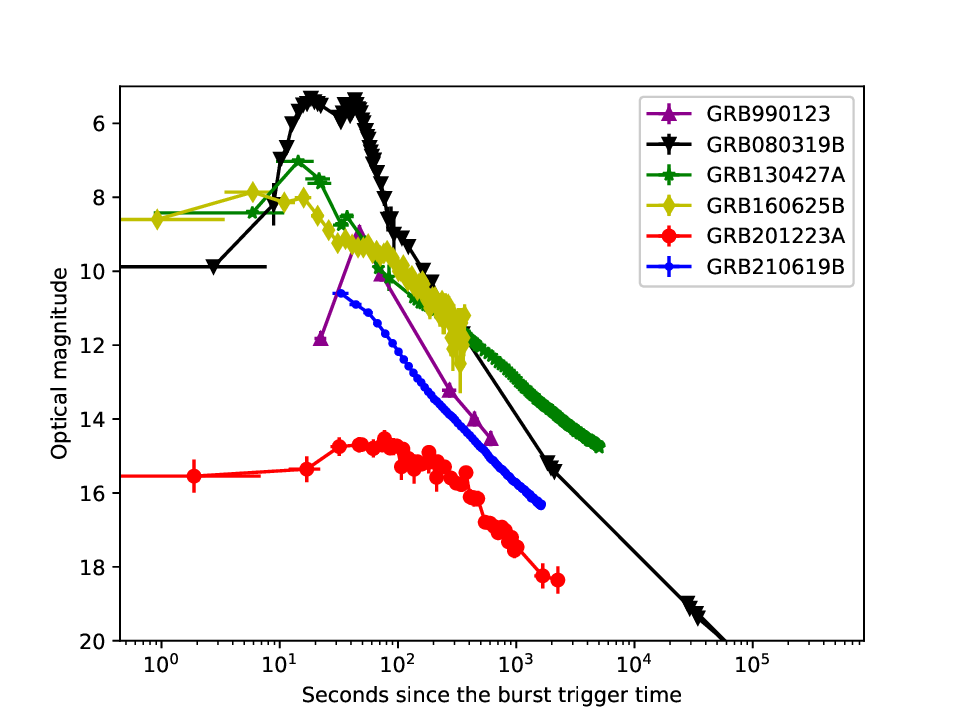}
   \caption{
   The optical light curve in the GRB sample  with extremely early optical emission, showing a wide distribution in the brightness at several seconds right after these bursts.} 
   \label{GRB_redshift}
   \end{figure}

On 2020 December 23, a complete prompt phase and the 
followed transition to the early afterglow
were detected in optical band for GRB\,201223A  by GWAC and F60A (\cite{2023NatAs...7..724X}), in which
the onset of optical and gamma-ray emissions was detected simultaneously. 
The optical prompt emission level predicted by an extrapolation of the gamma-ray emission
was found to be lower than the detected level by about four orders of magnitude, suggesting an additional contribution to the optical prompt emission in this GRB. 
Combing the prompt and afterglow phases recorded by GWAC and F60A respectively,
a detailed light-curve analysis shows that the behavior of the optical light curve 
is fully consistent with the predictions of a forward shock, rather than a reverse shock,
even during the very early phase. The forward shock scenario at such very early phase enables us to place a constraint 
on the mass of the progenitor, which suggests GRB\,201223A is likely produced by a massive Wolf-Rayet star with a stellar mass less than 3.8$M_\odot$. The comparison of optical light curves of GRB 201223A with GRB sample (\cite{1999Natur.398..400A,2008Natur.455..183R,2014Sci...343...38V,2017Natur.547..425T,2023NatAs...7..843O}) with extremely early optical emission were presented in Figure. \ref{GRB_redshift}, indicating the optical brightness during the prompt phase is distributed widely  over several orders of magnitude.

After the lunch of SVOM on  2024 June 22, GWAC has covered the sky for six GRBs triggered by\textit{SVOM} \citep{Wei_2016arXiv161006892W,Cordier+etal+2026a}, \textit{Einstein Probe}(EP) \citep{2022hxga.book...86Y} , 
\textit{Neil Gehrels Swift Observatory}
(Swift) and 
\textit{Fermi} \citep{2009ApJ...702..791M} during the prompt emissions. They are GRB 251025B \citep{2025GCN.42437....1H}, GRB 251122A\citep{2025GCN.42800....1L}, GRB 250925A \citep{2025GCN.42006....1B,2025GCN.42001....1Z}, GRB 250404A \citep{2025GCN.40050....1F,2025GCN.40051....1H}, EP251102a \citep{2025GCN.42548....1L}
and EP251124a \citep{2025GCN.42816....1Z},
thanks to the alert chain and observation planing\citep{Han+etal+2026}. Upper limits in optical have been obtained for all these GRBs. All the data are still under analysis.

\subsection{Superflares from late-type stars}

Highly energetic flares with total energies exceeding $10^{32}\ \mathrm{erg}$ and 
typical durations from minutes to hours have
been frequently reported for solar-type or late-type main-sequence stars across
multiple wavelengths from near infrared to X-ray. It is widely believed that 
they result from magnetic reconnection (e.g., \cite{2017NatAs...1E.184S}).
The study of stellar flares is important for exploring the underlying physics of magnetic activity,  
and for constraining the event onset by well-sampled light curves, and for investigating the energy distribution and magnetic background, and for detecting possible stellar coronal mass ejections (CMEs, e.g., \cite{2022SerAJ.205....1L}), that could impact extraterrestrial life theoretically either by tearing off most of the atmosphere of an exoplanet over a long timescale (e.g., \cite{2007AsBio...7..167K,2017ApJ...846...31C}), or by chemically generating greenhouse gas and Hydrogen cyanide on a short timescale (e.g.,  \cite{2016NatGe...9..452A}).

To date, about 200 superflares with amplitudes greater than 0.8 magnitudes in the $R-$band have been detected in real time by GWAC system, even though only part of the system was operational during this period. 
The characteristics of a large fraction of stars selected by GWAC flares in their quiescent state have been studies (\cite{2024ApJ...971..114L}). 
The photometric follow-up observations in multi-bands have 
been obtained rapidly for a large percentage of these events, among them two superflares from ultra-cool stars have amplitudes of about 10 mag in the $R-$band,
placing them among the most powerful white flares of ultra-cool stars (\cite{2021ApJ...909..106X,2024MNRAS.527.2232X}).
With rapid spectroscopic follow-ups by the NAOC 2.16m telescope (\cite{2016PASP..128k5005F}) at Xinglong 
observatory, four CME candidates have been identified due to their high velocity H$\alpha$ emission line wings (\cite{2020AJ....159...35W,2021ApJ...916...92W}).  
The maximum projected velocity is determined to be $\sim500-800\ \mathrm{km\ s^{-1}}$, and 
the ejected mass is estimated to be $\sim10^{18-19}$g.

Due to the limited computational resources, a large fraction of low-amplitude flares are in fact filtered out by the real-time detection pipeline, and can be picked up by offline pipelines with more robust algorithms.
For example, 
\cite{2023RAA....23a5016L} discovered additional 43 stellar flares by examining a GWAC training dataset offline.
A subsequent statistical study revealed that 34 stars of these flares have rotation or orbital periods less than 5.4 days and only one star has a long period of $\sim$10.42 days. Further studies (\cite{2023RAA....23a5016L}) suggested that 
some of the 43 GWAC flares may be energetic enough to destroy the ozone layer of the 
(if any) habitable exoplanets, and none could trigger periotic chemistry on the planets. 
In addition to the energetic stellar flares detected by the real-time pipeline, 
\cite{2023ApJ...954..142L} recently identified 27 flares in  archival GWAC data
from a very nearby M7 active star EI\,Cnc, located at only 5.2 pc from Earth. 
The sample yielded several interesting findings. On  one hand, the star is found to be less
active than it was 30 years ago. On the other hand, the sample enables us to extend 
the flare energy vs. effective duration relationship previously established 
for solar-type stars to M dwarfs, which suggests a common flare mechanism for 
stars ranging from  G to late M type. 

\subsection{Fast radio bursts}
Fast Radio Burst (FRB) is a very fast, lasting milliseconds to seconds, outburst in radio band  (\cite{2022hxga.book...28X}). From the perspective of observation, there are two types of FRBs: repeated FRBs and one-off FRBs.  Since the first report in 2007 (\cite{2007Sci...318..777L}), several hundred bright FRBs have been detected by the CHIME facility\footnote{https://www.chime-frb.ca/catalog}. The estimated frequency of FRBs is 1-10000 per day for the whole sky. During their active
periods, thousands of bursts have been, however, detected in some repeated FRBs by the intensive observations performed by FAST (\cite{2021Natur.598..267L}), which implies an enhanced FRB frequency 
up to a hundred FRBs per hour.

The nature of FRBs is still a mystery. Dozens of models have been proposed to interpret this puzzling phenomenon in past decade, although there is accumulating evidence supporting that  
at least a fraction of repeated FRBs are associated with young active magnetars (\cite{2021NatAs...5..378L}). Searching for the multi-wavelength counterparts associated to FRBs is essential for unraveling the nature of FRBs, including radiation mechanism, central engine, as well as the characteristics of their progenitors. Although many efforts have been made to search for such counterparts based on  various theoretical predictions, no credible counterparts have been found, except for 
the X-ray counterpart J1935+21, a Galactic magnetar (\cite{2021NatAs...5..378L}).  

Due to the high cadence, GWAC has a potential capability to search for the short lasting 
counterparts associated with FRBs by simultaneous observations in optical and radio bands. 
Such observation has been carried out by GWAC and FAST for FRB\,180330 ((\cite{2021ApJ...922...78X})), which provides a good constraint on the spectral 
index from optical to radio bands for this event. 
In the future, we plan to search for FRB's optical counterparts by GWAC through 
more planed simultaneous observations and an off-line search of bright optical transients with a timescale of seconds in the GWAC archive data.

\subsection{Dwarf nova}

WZ Sge-type objects are the SU UMa-type DNe, a subtype of DNe, 
with extremely long intervals between super outbursts that
are related to mass transfer (\cite{2015PASJ...67..108K}). The amplitudes of the super outbursts is typically
6–8 mag. Some WZ Sge-type DNe exhibit multiple 
rebrightenings after a dip (e.g., \cite{2002PASP..114..721P,2006PASJ...58L..23I,2015PASJ...67...52M}). 
WZ Sge-type DNe are believed to have a degenerate BD companion, 
and extremely short orbital periods. The discovered WZ Sge-type DNe are rare. The number is now much smaller than that predicted by current Milkway evolution model in which  
$\sim40\%–70\%$ of CVs should have a BD companion and
have passed the orbital period minimum, i.e., “period bouncer” (e.g., \cite{2018MNRAS.481.2523P}). 

One WZ sge-type dwarf nova (DN\footnote{A kind of Cataclysmic variables (CVs). CVs are semidetached binaries that consist of a white dwarf (the primary) and a low-mass star
(the secondary or donor) (\cite{1995cvs..book.....W}). The outbursts of CVs are believed to be resulted from 
a thermal instability in the accretion disk (e.g., \cite{1981A&A...104L..10M,1989PASJ...41.1005O,1996PASP..108...39O,2001NewAR..45..449L}) formed from matter lost by the donor 
(either a low-mass main-sequence star or a brown dwarf, BD) through the inner
Lagrangian point. }) was independently identified by GWAC in 2018 run, as an example, showing 
a long lasting (∼2 weeks) multiple rebrightenings with amplitudes of 3–4 mag.
The early superhump associated with a double-wave modulation and the extremely low mass 
ratio were found (\cite{2020AJ....159...35W}).

\section{Summary}

GWAC is a unique ground-based array of ultra-wide-angle cameras located at Xinglong Observatory, China, with a total FoV of 3600 square degrees and a cadence of 3 seconds, reaching 16 mag in visual band under a new Moon.
In this paper, we provide an overview of GWAC, including its scientific objectives, hardware and software, observing strategy and data processing.  Thanks to the high-cadence observations with large sky coverage, GWAC has the potential capability to study the bright optical transients on the timescale of seconds.  
These early science is also briefly summarized.  The prompt optical emission and the prompt to afterglow transition of a long duration gamma-ray burst have been detected. Hundreds of superflares from later-type dwarfs are self-triggered with fine temporal resolution, allowing the search for new components during the flares. Detailed analysis of superflares from a very nearby star, confirming the change in long-term stellar activity. Statistical studies for 43 superflares identified from a subset of the machine learning sample are also reported. Four flare-associated CME candidates were discovered by rapid follow-up with multi-wavelength photometries and spectroscopes. Additionally, constraints on the optical emission of fast radio bursts from synchronous observations with GWAC and FAST are presented. 

Besides,
the search for optical emission associated to 15 gravitational-wave events triggered by LIGO/VIRGO detector during O3 including two BNS events (S190425z\cite{2019GCN.24168....1L} and S190901ap\cite{2019GCN.25606....1L}) are also performed. However, no optical transient was detected for any LIGO-VIRGO events observed.  

\textit{SVOM} has launched at 2024 June 22. The joint simultaneous observations by GWAC with \textit{SVOM} is scheduled based on the observation plan for each. GWAC also could cover the sky which is observing by other missions, like \textit{Swift, EP or Fermi}. These joint observations are essential for exploring the radiation mechanism of the prompt emission from gamma-ray bursts triggered by \textit{SVOM}, as well as the characteristics of  the interactions between the relativistic outflow and the external medium, and the nature of the progenitors.


\normalem
\section{Acknowledgments}
The Space-based multi-band astronomical Variable Objects Monitor (SVOM) is a joint Chinese-French mission led by the Chinese National Space Administration (CNSA), the French Space Agency (CNES), and the Chinese Academy of Sciences (CAS). We gratefully acknowledge the unwavering support of NSSC, IAMCAS, XIOPM, NAOC, IHEP, CNES, CEA, and CNRS.
This work is supported by the Strategic Priority Research Program of the Chinese Academy of Sciences(Grant No.XDB0550401)，and by the National Natural Science Foundation of China (grant Nos. 12494571 and 12494570，12494573， 12133003).
The authors are thankful for support from the National Key R\&D Program of China (grant Nos. 2024YFA161170* and 2024YFA1611700). 


\section*{Author Contributions for GWAC system} 
Jiayan WEI is the principal investigator for both SVOM and GWAC, proposed the project and leading the development.
Liping XIN built the initial version of the pipeline for real-time detection of transient phenomena, and is in responsible for the scientific operations of GWAC.
Lei Huang is responsible for the building and control of the telescope, dome, and autofocus system.
Hongbo CAI is responsible for the algorithm and implementation of real-time data processing based on catalog differential method.
Xuhui HAN managed alert reception and observation scheduling.
Yang XU implemented the scientific database and automated follow-up observation system. He was in charge of developing the image subtraction-based transient detection pipeline and was responsible for the operational status monitoring system.
Yulei QIU is responsible for the implementation of the astrometric calibration software.
Huali Li handled refined data processing.
Xiaomeng LU and Rongsong ZHANG are responsible for detector control and image acquisition.
Pinpin Zhang and Yujie XIAO are responsible for real-time system operation scheduling and anomaly handling, and they also participate in observation scheduling.
Zigao DAI and Xiangyu WANG are one of the funding sponsors for GWAC and participate in scientific operations.
Enwei LIANG and Xianggao WANG are the funding sponsors for the F60A/B follow-up telescopes and participate in GWAC scientific operations.
Yuangui YANG is the funding sponsor for the F30/F50 follow-up telescopes and participates in GWAC scientific operations.
Jingran XU and Yinuo MA participate in data processing and transient identification.
Jing WANG participate in the scientific operations and led the science on stellar flares.
Guangwei LI, Jingsong DENG, Dangwei XU, Chao WU, and Zhuheng YAO participate in the scientific operations.
Yongtong ZHENG and Wenlong DONG are responsible for observational operations.

\bibliographystyle{raa}
\bibliography{bibtex}

@article{2020MNRAS.497..726G,
	adsurl = {https://ui.adsabs.harvard.edu/abs/2020MNRAS.497..726G},
	archiveprefix = {arXiv},
	author = {{Gompertz}, B.~P. and {Cutter}, R. and {Steeghs}, D. and {Galloway}, D.~K. and {Lyman}, J. and {Ulaczyk}, K. and {Dyer}, M.~J. and {Ackley}, K. and {Dhillon}, V.~S. and {O'Brien}, P.~T. and {Ramsay}, G. and {Poshyachinda}, S. and {Kotak}, R. and {Nuttall}, L. and {Breton}, R.~P. and {Pall{\'e}}, E. and {Pollacco}, D. and {Thrane}, E. and {Aukkaravittayapun}, S. and {Awiphan}, S. and {Brown}, M.~J.~I. and {Burhanudin}, U. and {Chote}, P. and {Chrimes}, A.~A. and {Daw}, E. and {Duffy}, C. and {Eyles-Ferris}, R.~A.~J. and {Heikkil{\"a}}, T. and {Irawati}, P. and {Kennedy}, M.~R. and {Killestein}, T. and {Levan}, A.~J. and {Littlefair}, S. and {Makrygianni}, L. and {Marsh}, T. and {Mata S{\'a}nchez}, D. and {Mattila}, S. and {Maund}, J. and {McCormac}, J. and {Mkrtichian}, D. and {Mong}, Y.-L. and {Mullaney}, J. and {M{\"u}ller}, B. and {Obradovic}, A. and {Rol}, E. and {Sawangwit}, U. and {Stanway}, E.~R. and {Starling}, R.~L.~C. and {Str{\o}m}, P.~A. and {Tooke}, S. and {West}, R. and {Wiersema}, K.},
	doi = {10.1093/mnras/staa1845},
	eprint = {2004.00025},
	journal = {\mnras},
	month = sep,
	number = {1},
	pages = {726-738},
	primaryclass = {astro-ph.HE},
	title = {{Searching for electromagnetic counterparts to gravitational-wave merger events with the prototype Gravitational-Wave Optical Transient Observer (GOTO-4)}},
	volume = {497},
	year = 2020}

@article{2019ApJ...873..111I,
	adsurl = {https://ui.adsabs.harvard.edu/abs/2019ApJ...873..111I},
	archiveprefix = {arXiv},
	author = {{Ivezi{\'c}}, {\v{Z}}eljko and {Kahn}, Steven M. and {Tyson}, J. Anthony and {Abel}, Bob and {Acosta}, Emily and {Allsman}, Robyn and {Alonso}, David and {AlSayyad}, Yusra and {Anderson}, Scott F. and {Andrew}, John and {Angel}, James Roger P. and {Angeli}, George Z. and {Ansari}, Reza and {Antilogus}, Pierre and {Araujo}, Constanza and {Armstrong}, Robert and {Arndt}, Kirk T. and {Astier}, Pierre and {Aubourg}, {\'E}ric and {Auza}, Nicole and {Axelrod}, Tim S. and {Bard}, Deborah J. and {Barr}, Jeff D. and {Barrau}, Aurelian and {Bartlett}, James G. and {Bauer}, Amanda E. and {Bauman}, Brian J. and {Baumont}, Sylvain and {Bechtol}, Ellen and {Bechtol}, Keith and {Becker}, Andrew C. and {Becla}, Jacek and {Beldica}, Cristina and {Bellavia}, Steve and {Bianco}, Federica B. and {Biswas}, Rahul and {Blanc}, Guillaume and {Blazek}, Jonathan and {Blandford}, Roger D. and {Bloom}, Josh S. and {Bogart}, Joanne and {Bond}, Tim W. and {Booth}, Michael T. and {Borgland}, Anders W. and {Borne}, Kirk and {Bosch}, James F. and {Boutigny}, Dominique and {Brackett}, Craig A. and {Bradshaw}, Andrew and {Brandt}, William Nielsen and {Brown}, Michael E. and {Bullock}, James S. and {Burchat}, Patricia and {Burke}, David L. and {Cagnoli}, Gianpietro and {Calabrese}, Daniel and {Callahan}, Shawn and {Callen}, Alice L. and {Carlin}, Jeffrey L. and {Carlson}, Erin L. and {Chandrasekharan}, Srinivasan and {Charles-Emerson}, Glenaver and {Chesley}, Steve and {Cheu}, Elliott C. and {Chiang}, Hsin-Fang and {Chiang}, James and {Chirino}, Carol and {Chow}, Derek and {Ciardi}, David R. and {Claver}, Charles F. and {Cohen-Tanugi}, Johann and {Cockrum}, Joseph J. and {Coles}, Rebecca and {Connolly}, Andrew J. and {Cook}, Kem H. and {Cooray}, Asantha and {Covey}, Kevin R. and {Cribbs}, Chris and {Cui}, Wei and {Cutri}, Roc and {Daly}, Philip N. and {Daniel}, Scott F. and {Daruich}, Felipe and {Daubard}, Guillaume and {Daues}, Greg and {Dawson}, William and {Delgado}, Francisco and {Dellapenna}, Alfred and {de Peyster}, Robert and {de Val-Borro}, Miguel and {Digel}, Seth W. and {Doherty}, Peter and {Dubois}, Richard and {Dubois-Felsmann}, Gregory P. and {Durech}, Josef and {Economou}, Frossie and {Eifler}, Tim and {Eracleous}, Michael and {Emmons}, Benjamin L. and {Fausti Neto}, Angelo and {Ferguson}, Henry and {Figueroa}, Enrique and {Fisher-Levine}, Merlin and {Focke}, Warren and {Foss}, Michael D. and {Frank}, James and {Freemon}, Michael D. and {Gangler}, Emmanuel and {Gawiser}, Eric and {Geary}, John C. and {Gee}, Perry and {Geha}, Marla and {Gessner}, Charles J.~B. and {Gibson}, Robert R. and {Gilmore}, D. Kirk and {Glanzman}, Thomas and {Glick}, William and {Goldina}, Tatiana and {Goldstein}, Daniel A. and {Goodenow}, Iain and {Graham}, Melissa L. and {Gressler}, William J. and {Gris}, Philippe and {Guy}, Leanne P. and {Guyonnet}, Augustin and {Haller}, Gunther and {Harris}, Ron and {Hascall}, Patrick A. and {Haupt}, Justine and {Hernandez}, Fabio and {Herrmann}, Sven and {Hileman}, Edward and {Hoblitt}, Joshua and {Hodgson}, John A. and {Hogan}, Craig and {Howard}, James D. and {Huang}, Dajun and {Huffer}, Michael E. and {Ingraham}, Patrick and {Innes}, Walter R. and {Jacoby}, Suzanne H. and {Jain}, Bhuvnesh and {Jammes}, Fabrice and {Jee}, M. James and {Jenness}, Tim and {Jernigan}, Garrett and {Jevremovi{\'c}}, Darko and {Johns}, Kenneth and {Johnson}, Anthony S. and {Johnson}, Margaret W.~G. and {Jones}, R. Lynne and {Juramy-Gilles}, Claire and {Juri{\'c}}, Mario and {Kalirai}, Jason S. and {Kallivayalil}, Nitya J. and {Kalmbach}, Bryce and {Kantor}, Jeffrey P. and {Karst}, Pierre and {Kasliwal}, Mansi M. and {Kelly}, Heather and {Kessler}, Richard and {Kinnison}, Veronica and {Kirkby}, David and {Knox}, Lloyd and {Kotov}, Ivan V. and {Krabbendam}, Victor L. and {Krughoff}, K. Simon and {Kub{\'a}nek}, Petr and {Kuczewski}, John and {Kulkarni}, Shri and {Ku}, John and {Kurita}, Nadine R. and {Lage}, Craig S. and {Lambert}, Ron and {Lange}, Travis and {Langton}, J. Brian and {Le Guillou}, Laurent and {Levine}, Deborah and {Liang}, Ming and {Lim}, Kian-Tat and {Lintott}, Chris J. and {Long}, Kevin E. and {Lopez}, Margaux and {Lotz}, Paul J. and {Lupton}, Robert H. and {Lust}, Nate B. and {MacArthur}, Lauren A. and {Mahabal}, Ashish and {Mandelbaum}, Rachel and {Markiewicz}, Thomas W. and {Marsh}, Darren S. and {Marshall}, Philip J. and {Marshall}, Stuart and {May}, Morgan and {McKercher}, Robert and {McQueen}, Michelle and {Meyers}, Joshua and {Migliore}, Myriam and {Miller}, Michelle and {Mills}, David J.},
	doi = {10.3847/1538-4357/ab042c},
	eid = {111},
	eprint = {0805.2366},
	journal = {\apj},
	month = mar,
	number = {2},
	pages = {111},
	primaryclass = {astro-ph},
	title = {{LSST: From Science Drivers to Reference Design and Anticipated Data Products}},
	volume = {873},
	year = 2019}

@article{2019PASP..131a8002B,
	adsurl = {https://ui.adsabs.harvard.edu/abs/2019PASP..131a8002B},
	archiveprefix = {arXiv},
	author = {{Bellm}, Eric C. and {Kulkarni}, Shrinivas R. and {Graham}, Matthew J. and {Dekany}, Richard and {Smith}, Roger M. and {Riddle}, Reed and {Masci}, Frank J. and {Helou}, George and {Prince}, Thomas A. and {Adams}, Scott M. and {Barbarino}, C. and {Barlow}, Tom and {Bauer}, James and {Beck}, Ron and {Belicki}, Justin and {Biswas}, Rahul and {Blagorodnova}, Nadejda and {Bodewits}, Dennis and {Bolin}, Bryce and {Brinnel}, Valery and {Brooke}, Tim and {Bue}, Brian and {Bulla}, Mattia and {Burruss}, Rick and {Cenko}, S. Bradley and {Chang}, Chan-Kao and {Connolly}, Andrew and {Coughlin}, Michael and {Cromer}, John and {Cunningham}, Virginia and {De}, Kishalay and {Delacroix}, Alex and {Desai}, Vandana and {Duev}, Dmitry A. and {Eadie}, Gwendolyn and {Farnham}, Tony L. and {Feeney}, Michael and {Feindt}, Ulrich and {Flynn}, David and {Franckowiak}, Anna and {Frederick}, S. and {Fremling}, C. and {Gal-Yam}, Avishay and {Gezari}, Suvi and {Giomi}, Matteo and {Goldstein}, Daniel A. and {Golkhou}, V. Zach and {Goobar}, Ariel and {Groom}, Steven and {Hacopians}, Eugean and {Hale}, David and {Henning}, John and {Ho}, Anna Y.~Q. and {Hover}, David and {Howell}, Justin and {Hung}, Tiara and {Huppenkothen}, Daniela and {Imel}, David and {Ip}, Wing-Huen and {Ivezi{\'c}}, {\v{Z}}eljko and {Jackson}, Edward and {Jones}, Lynne and {Juric}, Mario and {Kasliwal}, Mansi M. and {Kaspi}, S. and {Kaye}, Stephen and {Kelley}, Michael S.~P. and {Kowalski}, Marek and {Kramer}, Emily and {Kupfer}, Thomas and {Landry}, Walter and {Laher}, Russ R. and {Lee}, Chien-De and {Lin}, Hsing Wen and {Lin}, Zhong-Yi and {Lunnan}, Ragnhild and {Giomi}, Matteo and {Mahabal}, Ashish and {Mao}, Peter and {Miller}, Adam A. and {Monkewitz}, Serge and {Murphy}, Patrick and {Ngeow}, Chow-Choong and {Nordin}, Jakob and {Nugent}, Peter and {Ofek}, Eran and {Patterson}, Maria T. and {Penprase}, Bryan and {Porter}, Michael and {Rauch}, Ludwig and {Rebbapragada}, Umaa and {Reiley}, Dan and {Rigault}, Mickael and {Rodriguez}, Hector and {van Roestel}, Jan and {Rusholme}, Ben and {van Santen}, Jakob and {Schulze}, S. and {Shupe}, David L. and {Singer}, Leo P. and {Soumagnac}, Maayane T. and {Stein}, Robert and {Surace}, Jason and {Sollerman}, Jesper and {Szkody}, Paula and {Taddia}, F. and {Terek}, Scott and {Van Sistine}, Angela and {van Velzen}, Sjoert and {Vestrand}, W. Thomas and {Walters}, Richard and {Ward}, Charlotte and {Ye}, Quan-Zhi and {Yu}, Po-Chieh and {Yan}, Lin and {Zolkower}, Jeffry},
	doi = {10.1088/1538-3873/aaecbe},
	eprint = {1902.01932},
	journal = {\pasp},
	month = jan,
	number = {995},
	pages = {018002},
	primaryclass = {astro-ph.IM},
	title = {{The Zwicky Transient Facility: System Overview, Performance, and First Results}},
	volume = {131},
	year = 2019}

@inproceedings{2014AAS...22323603S,
	adsurl = {https://ui.adsabs.harvard.edu/abs/2014AAS...22323603S},
	author = {{Shappee}, Benjamin and {Prieto}, J. and {Stanek}, K.~Z. and {Kochanek}, C.~S. and {Holoien}, T. and {Jencson}, J. and {Basu}, U. and {Beacom}, J.~F. and {Szczygiel}, D. and {Pojmanski}, G. and {Brimacombe}, J. and {Dubberley}, M. and {Elphick}, M. and {Foale}, S. and {Hawkins}, E. and {Mullins}, D. and {Rosing}, W. and {Ross}, R. and {Walker}, Z.},
	booktitle = {American Astronomical Society Meeting Abstracts \#223},
	eid = {236.03},
	month = jan,
	pages = {236.03},
	series = {American Astronomical Society Meeting Abstracts},
	title = {{All Sky Automated Survey for SuperNovae (ASAS-SN or ``Assassin'')}},
	volume = {223},
	year = 2014}

@article{2011PASP..123...58T,
	adsurl = {https://ui.adsabs.harvard.edu/abs/2011PASP..123...58T},
	archiveprefix = {arXiv},
	author = {{Tonry}, John L.},
	doi = {10.1086/657997},
	eprint = {1011.1028},
	journal = {\pasp},
	month = jan,
	number = {899},
	pages = {58},
	primaryclass = {astro-ph.IM},
	title = {{An Early Warning System for Asteroid Impact}},
	volume = {123},
	year = 2011}

@article{2010AdAst2010E..30L,
	adsurl = {https://ui.adsabs.harvard.edu/abs/2010AdAst2010E..30L},
	archiveprefix = {arXiv},
	author = {{Lipunov}, Vladimir and {Kornilov}, Victor and {Gorbovskoy}, Evgeny and {Shatskij}, Nikolaj and {Kuvshinov}, Dmitry and {Tyurina}, Nataly and {Belinski}, Alexander and {Krylov}, Alexander and {Balanutsa}, Pavel and {Chazov}, Vadim and {Kuznetsov}, Artem and {Kortunov}, Petr and {Sankovich}, Anatoly and {Tlatov}, Andrey and {Parkhomenko}, A. and {Krushinsky}, Vadim and {Zalozhnyh}, Ivan and {Popov}, A. and {Kopytova}, Taisia and {Ivanov}, Kirill and {Yazev}, Sergey and {Yurkov}, Vladimir},
	doi = {10.1155/2010/349171},
	eid = {349171},
	eprint = {0907.0827},
	journal = {Advances in Astronomy},
	month = jan,
	pages = {349171},
	primaryclass = {astro-ph.HE},
	title = {{Master Robotic Net}},
	volume = {2010},
	year = 2010}

@article{2023NatAs...7..724X,
	adsurl = {https://ui.adsabs.harvard.edu/abs/2023NatAs...7..724X},
	archiveprefix = {arXiv},
	author = {{Xin}, Liping and {Han}, Xuhui and {Li}, Huali and {Zhang}, Bing and {Wang}, Jing and {Turpin}, Damien and {Yang}, Xing and {Qiu}, Yulei and {Liang}, Enwei and {Dai}, Zigao and {Cai}, Hongbo and {Lu}, Xiaomeng and {Wang}, Xiang-Yu and {Huang}, Lei and {Wang}, Xianggao and {Wu}, Chao and {Gao}, He and {Ren}, Jia and {Zhang}, Lulu and {Yang}, Yuangui and {Deng}, Jingsong and {Wei}, Jianyan},
	doi = {10.1038/s41550-023-01930-0},
	eprint = {2304.04669},
	journal = {Nature Astronomy},
	month = jun,
	pages = {724-730},
	primaryclass = {astro-ph.HE},
	title = {{Prompt-to-afterglow transition of optical emission in a long gamma-ray burst consistent with a fireball}},
	volume = {7},
	year = 2023}

@article{2019A&A...628A..59O,
	adsurl = {https://ui.adsabs.harvard.edu/abs/2019A&A...628A..59O},
	archiveprefix = {arXiv},
	author = {{Oganesyan}, G. and {Nava}, L. and {Ghirlanda}, G. and {Melandri}, A. and {Celotti}, A.},
	doi = {10.1051/0004-6361/201935766},
	eid = {A59},
	eprint = {1904.11086},
	journal = {\aap},
	month = aug,
	pages = {A59},
	primaryclass = {astro-ph.HE},
	title = {{Prompt optical emission as a signature of synchrotron radiation in gamma-ray bursts}},
	volume = {628},
	year = 2019}

@article{2015PhR...561....1K,
	adsurl = {https://ui.adsabs.harvard.edu/abs/2015PhR...561....1K},
	archiveprefix = {arXiv},
	author = {{Kumar}, Pawan and {Zhang}, Bing},
	doi = {10.1016/j.physrep.2014.09.008},
	eprint = {1410.0679},
	journal = {\physrep},
	month = feb,
	pages = {1-109},
	primaryclass = {astro-ph.HE},
	title = {{The physics of gamma-ray bursts \& relativistic jets}},
	volume = {561},
	year = 2015}

@article{2014Sci...343...38V,
	adsurl = {https://ui.adsabs.harvard.edu/abs/2014Sci...343...38V},
	archiveprefix = {arXiv},
	author = {{Vestrand}, W.~T. and {Wren}, J.~A. and {Panaitescu}, A. and {Wozniak}, P.~R. and {Davis}, H. and {Palmer}, D.~M. and {Vianello}, G. and {Omodei}, N. and {Xiong}, S. and {Briggs}, M.~S. and {Elphick}, M. and {Paciesas}, W. and {Rosing}, W.},
	doi = {10.1126/science.1242316},
	eprint = {1311.5489},
	journal = {Science},
	month = jan,
	number = {6166},
	pages = {38-41},
	primaryclass = {astro-ph.HE},
	title = {{The Bright Optical Flash and Afterglow from the Gamma-Ray Burst GRB 130427A}},
	volume = {343},
	year = 2014}

@article{2008Natur.455..183R,
	adsurl = {https://ui.adsabs.harvard.edu/abs/2008Natur.455..183R},
	archiveprefix = {arXiv},
	author = {{Racusin}, J.~L. and {Karpov}, S.~V. and {Sokolowski}, M. and {Granot}, J. and {Wu}, X.~F. and {Pal'Shin}, V. and {Covino}, S. and {van der Horst}, A.~J. and {Oates}, S.~R. and {Schady}, P. and {Smith}, R.~J. and {Cummings}, J. and {Starling}, R.~L.~C. and {Piotrowski}, L.~W. and {Zhang}, B. and {Evans}, P.~A. and {Holland}, S.~T. and {Malek}, K. and {Page}, M.~T. and {Vetere}, L. and {Margutti}, R. and {Guidorzi}, C. and {Kamble}, A.~P. and {Curran}, P.~A. and {Beardmore}, A. and {Kouveliotou}, C. and {Mankiewicz}, L. and {Melandri}, A. and {O'Brien}, P.~T. and {Page}, K.~L. and {Piran}, T. and {Tanvir}, N.~R. and {Wrochna}, G. and {Aptekar}, R.~L. and {Barthelmy}, S. and {Bartolini}, C. and {Beskin}, G.~M. and {Bondar}, S. and {Bremer}, M. and {Campana}, S. and {Castro-Tirado}, A. and {Cucchiara}, A. and {Cwiok}, M. and {D'Avanzo}, P. and {D'Elia}, V. and {Della Valle}, M. and {de Ugarte Postigo}, A. and {Dominik}, W. and {Falcone}, A. and {Fiore}, F. and {Fox}, D.~B. and {Frederiks}, D.~D. and {Fruchter}, A.~S. and {Fugazza}, D. and {Garrett}, M.~A. and {Gehrels}, N. and {Golenetskii}, S. and {Gomboc}, A. and {Gorosabel}, J. and {Greco}, G. and {Guarnieri}, A. and {Immler}, S. and {Jelinek}, M. and {Kasprowicz}, G. and {La Parola}, V. and {Levan}, A.~J. and {Mangano}, V. and {Mazets}, E.~P. and {Molinari}, E. and {Moretti}, A. and {Nawrocki}, K. and {Oleynik}, P.~P. and {Osborne}, J.~P. and {Pagani}, C. and {Pandey}, S.~B. and {Paragi}, Z. and {Perri}, M. and {Piccioni}, A. and {Ramirez-Ruiz}, E. and {Roming}, P.~W.~A. and {Steele}, I.~A. and {Strom}, R.~G. and {Testa}, V. and {Tosti}, G. and {Ulanov}, M.~V. and {Wiersema}, K. and {Wijers}, R.~A.~M.~J. and {Winters}, J.~M. and {Zarnecki}, A.~F. and {Zerbi}, F. and {M{\'e}sz{\'a}ros}, P. and {Chincarini}, G. and {Burrows}, D.~N.},
	doi = {10.1038/nature07270},
	eprint = {0805.1557},
	journal = {\nat},
	month = sep,
	number = {7210},
	pages = {183-188},
	primaryclass = {astro-ph},
	title = {{Broadband observations of the naked-eye {\ensuremath{\gamma}}-ray burst GRB080319B}},
	volume = {455},
	year = 2008}

@article{2007ChJAA...7....1Z,
	adsurl = {https://ui.adsabs.harvard.edu/abs/2007ChJAA...7....1Z},
	archiveprefix = {arXiv},
	author = {{Zhang}, Bing},
	doi = {10.1088/1009-9271/7/1/01},
	eprint = {astro-ph/0701520},
	journal = {\cjaa},
	month = feb,
	number = {1},
	pages = {1-50},
	primaryclass = {astro-ph},
	title = {{Gamma-Ray Bursts in the Swift Era}},
	volume = {7},
	year = 2007}

@article{2006Natur.442..172V,
	adsurl = {https://ui.adsabs.harvard.edu/abs/2006Natur.442..172V},
	archiveprefix = {arXiv},
	author = {{Vestrand}, W.~T. and {Wren}, J.~A. and {Wozniak}, P.~R. and {Aptekar}, R. and {Golentskii}, S. and {Pal'Shin}, V. and {Sakamoto}, T. and {White}, R.~R. and {Evans}, S. and {Casperson}, D. and {Fenimore}, E.},
	doi = {10.1038/nature04913},
	eprint = {astro-ph/0605472},
	journal = {\nat},
	month = jul,
	number = {7099},
	pages = {172-175},
	primaryclass = {astro-ph},
	title = {{Energy input and response from prompt and early optical afterglow emission in {\ensuremath{\gamma}}-ray bursts}},
	volume = {442},
	year = 2006}

@article{2004RvMP...76.1143P,
	adsurl = {https://ui.adsabs.harvard.edu/abs/2004RvMP...76.1143P},
	archiveprefix = {arXiv},
	author = {{Piran}, Tsvi},
	doi = {10.1103/RevModPhys.76.1143},
	eprint = {astro-ph/0405503},
	journal = {Reviews of Modern Physics},
	month = oct,
	number = {4},
	pages = {1143-1210},
	primaryclass = {astro-ph},
	title = {{The physics of gamma-ray bursts}},
	volume = {76},
	year = 2004}

@article{Wei_2016arXiv161006892W,
	adsurl = {https://ui.adsabs.harvard.edu/abs/2016arXiv161006892W},
	archiveprefix = {arXiv},
	author = {{Wei}, J. and {Cordier}, B. and {Antier}, S. and {Antilogus}, P. and {Atteia}, J. -L. and {Bajat}, A. and {Basa}, S. and {Beckmann}, V. and {Bernardini}, M.~G. and {Boissier}, S. and {Bouchet}, L. and {Burwitz}, V. and {Claret}, A. and {Dai}, Z. -G. and {Daigne}, F. and {Deng}, J. and {Dornic}, D. and {Feng}, H. and {Foglizzo}, T. and {Gao}, H. and {Gehrels}, N. and {Godet}, O. and {Goldwurm}, A. and {Gonzalez}, F. and {Gosset}, L. and {G{\"o}tz}, D. and {Gouiffes}, C. and {Grise}, F. and {Gros}, A. and {Guilet}, J. and {Han}, X. and {Huang}, M. and {Huang}, Y. -F. and {Jouret}, M. and {Klotz}, A. and {La Marle}, O. and {Lachaud}, C. and {Le Floch}, E. and {Lee}, W. and {Leroy}, N. and {Li}, L. -X. and {Li}, S.~C. and {Li}, Z. and {Liang}, E. -W. and {Lyu}, H. and {Mercier}, K. and {Migliori}, G. and {Mochkovitch}, R. and {O'Brien}, P. and {Osborne}, J. and {Paul}, J. and {Perinati}, E. and {Petitjean}, P. and {Piron}, F. and {Qiu}, Y. and {Rau}, A. and {Rodriguez}, J. and {Schanne}, S. and {Tanvir}, N. and {Vangioni}, E. and {Vergani}, S. and {Wang}, F. -Y. and {Wang}, J. and {Wang}, X. -G. and {Wang}, X. -Y. and {Watson}, A. and {Webb}, N. and {Wei}, J.~J. and {Willingale}, R. and {Wu}, C. and {Wu}, X. -F. and {Xin}, L. -P. and {Xu}, D. and {Yu}, S. and {Yu}, W. -F. and {Yu}, Y. -W. and {Zhang}, B. and {Zhang}, S. -N. and {Zhang}, Y. and {Zhou}, X.~L.},
	doi = {10.48550/arXiv.1610.06892},
	eid = {arXiv:1610.06892},
	eprint = {1610.06892},
	journal = {arXiv e-prints},
	month = oct,
	pages = {arXiv:1610.06892},
	primaryclass = {astro-ph.IM},
	title = {{The Deep and Transient Universe in the SVOM Era: New Challenges and Opportunities - Scientific prospects of the SVOM mission}},
	year = 2016}

@article{Xu_2020arXiv200300209X,
	adsurl = {https://ui.adsabs.harvard.edu/abs/2020arXiv200300209X},
	archiveprefix = {arXiv},
	author = {{Xu}, Yang and {Wang}, Jing and {Huang}, Maohai and {Wei}, Jianyan},
	doi = {10.48550/arXiv.2003.00209},
	eid = {arXiv:2003.00209},
	eprint = {2003.00209},
	journal = {arXiv e-prints},
	month = feb,
	pages = {arXiv:2003.00209},
	primaryclass = {astro-ph.IM},
	title = {{An algorithm of selection of meteor candidates in GWAC system}},
	year = 2020}

@article{1996A&AS..117..393B,
	adsurl = {https://ui.adsabs.harvard.edu/abs/1996A&AS..117..393B},
	author = {{Bertin}, E. and {Arnouts}, S.},
	doi = {10.1051/aas:1996164},
	journal = {\aaps},
	month = jun,
	pages = {393-404},
	title = {{SExtractor: Software for source extraction.}},
	volume = {117},
	year = 1996}

@article{2013AJ....145...44Z,
	adsurl = {https://ui.adsabs.harvard.edu/abs/2013AJ....145...44Z},
	archiveprefix = {arXiv},
	author = {{Zacharias}, N. and {Finch}, C.~T. and {Girard}, T.~M. and {Henden}, A. and {Bartlett}, J.~L. and {Monet}, D.~G. and {Zacharias}, M.~I.},
	doi = {10.1088/0004-6256/145/2/44},
	eid = {44},
	eprint = {1212.6182},
	journal = {\aj},
	month = feb,
	number = {2},
	pages = {44},
	primaryclass = {astro-ph.IM},
	title = {{The Fourth US Naval Observatory CCD Astrograph Catalog (UCAC4)}},
	volume = {145},
	year = 2013}

@article{2024MNRAS.527.2232X,
	adsurl = {https://ui.adsabs.harvard.edu/abs/2024MNRAS.527.2232X},
	archiveprefix = {arXiv},
	author = {{Xin}, Li-Ping and {Li}, Hua-li and {Wang}, Jing and {Han}, Xu-Hui and {Cai}, Hong-Bo and {Huang}, Xin-Bo and {Cao}, Jia-Xin and {Zhu}, Yi-Nan and {Wang}, Xiang-Gao and {Li}, Guang-Wei and {Ren}, Bin and {Gao}, Cheng and {Song}, Da and {Huang}, Lei and {Lu}, Xiao-Meng and {Bai}, Jian-Ying and {Qiu}, Yu-Lei and {Liang}, En-Wei and {Dai}, Zi-Gao and {Wang}, Xiang-Yu and {Wu}, Chao and {Deng}, Jing-Song and {Yang}, Yuan-Gui and {Wei}, Jian-Yan},
	doi = {10.1093/mnras/stad960},
	eprint = {2303.17415},
	journal = {\mnras},
	month = jan,
	number = {2},
	pages = {2232-2239},
	primaryclass = {astro-ph.SR},
	title = {{A huge-amplitude white-light superflare on a L0 brown dwarf discovered by GWAC survey}},
	volume = {527},
	year = 2024}

@article{2023ApJ...954..142L,
	adsurl = {https://ui.adsabs.harvard.edu/abs/2023ApJ...954..142L},
	archiveprefix = {arXiv},
	author = {{Li}, Hua-Li and {Wang}, Jing and {Xin}, Li-Ping and {Bai}, Jian-Ying and {Han}, Xu-Hui and {Cai}, Hong-Bo and {Huang}, Lei and {Lu}, Xiao-Meng and {Qiu}, Yu-Lei and {Wu}, Chao and {Li}, Guang-Wei and {Deng}, Jing-Song and {Xu}, Da-Wei and {Yang}, Yuan-Gui and {Wang}, Xiang-Gao and {Liang}, En-Wei and {Wei}, Jian-Yan},
	doi = {10.3847/1538-4357/ace59b},
	eid = {142},
	eprint = {2307.14594},
	journal = {\apj},
	month = sep,
	number = {2},
	pages = {142},
	primaryclass = {astro-ph.SR},
	title = {{White-light Superflare and Long-term Activity of the Nearby M7-type Binary EI Cnc Observed with GWAC System}},
	volume = {954},
	year = 2023}

@article{1973ApJ...182L..85K,
	adsurl = {https://ui.adsabs.harvard.edu/abs/1973ApJ...182L..85K},
	author = {{Klebesadel}, Ray W. and {Strong}, Ian B. and {Olson}, Roy A.},
	doi = {10.1086/181225},
	journal = {\apjl},
	month = jun,
	pages = {L85},
	title = {{Observations of Gamma-Ray Bursts of Cosmic Origin}},
	volume = {182},
	year = 1973}

@article{2004ApJ...611.1005G,
	adsurl = {https://ui.adsabs.harvard.edu/abs/2004ApJ...611.1005G},
	archiveprefix = {arXiv},
	author = {{Gehrels}, N. and {Chincarini}, G. and {Giommi}, P. and {Mason}, K.~O. and {Nousek}, J.~A. and {Wells}, A.~A. and {White}, N.~E. and {Barthelmy}, S.~D. and {Burrows}, D.~N. and {Cominsky}, L.~R. and {Hurley}, K.~C. and {Marshall}, F.~E. and {M{\'e}sz{\'a}ros}, P. and {Roming}, P.~W.~A. and {Angelini}, L. and {Barbier}, L.~M. and {Belloni}, T. and {Campana}, S. and {Caraveo}, P.~A. and {Chester}, M.~M. and {Citterio}, O. and {Cline}, T.~L. and {Cropper}, M.~S. and {Cummings}, J.~R. and {Dean}, A.~J. and {Feigelson}, E.~D. and {Fenimore}, E.~E. and {Frail}, D.~A. and {Fruchter}, A.~S. and {Garmire}, G.~P. and {Gendreau}, K. and {Ghisellini}, G. and {Greiner}, J. and {Hill}, J.~E. and {Hunsberger}, S.~D. and {Krimm}, H.~A. and {Kulkarni}, S.~R. and {Kumar}, P. and {Lebrun}, F. and {Lloyd-Ronning}, N.~M. and {Markwardt}, C.~B. and {Mattson}, B.~J. and {Mushotzky}, R.~F. and {Norris}, J.~P. and {Osborne}, J. and {Paczynski}, B. and {Palmer}, D.~M. and {Park}, H.-S. and {Parsons}, A.~M. and {Paul}, J. and {Rees}, M.~J. and {Reynolds}, C.~S. and {Rhoads}, J.~E. and {Sasseen}, T.~P. and {Schaefer}, B.~E. and {Short}, A.~T. and {Smale}, A.~P. and {Smith}, I.~A. and {Stella}, L. and {Tagliaferri}, G. and {Takahashi}, T. and {Tashiro}, M. and {Townsley}, L.~K. and {Tueller}, J. and {Turner}, M.~J.~L. and {Vietri}, M. and {Voges}, W. and {Ward}, M.~J. and {Willingale}, R. and {Zerbi}, F.~M. and {Zhang}, W.~W.},
	doi = {10.1086/422091},
	eprint = {astro-ph/0405233},
	journal = {\apj},
	month = aug,
	number = {2},
	pages = {1005-1020},
	primaryclass = {astro-ph},
	title = {{The Swift Gamma-Ray Burst Mission}},
	volume = {611},
	year = 2004}

@article{2005Natur.435..178V,
	adsurl = {https://ui.adsabs.harvard.edu/abs/2005Natur.435..178V},
	archiveprefix = {arXiv},
	author = {{Vestrand}, W.~T. and {Wozniak}, P.~R. and {Wren}, J.~A. and {Fenimore}, E.~E. and {Sakamoto}, T. and {White}, R.~R. and {Casperson}, D. and {Davis}, H. and {Evans}, S. and {Galassi}, M. and {McGowan}, K.~E. and {Schier}, J.~A. and {Asa}, J.~W. and {Barthelmy}, S.~D. and {Cummings}, J.~R. and {Gehrels}, N. and {Hullinger}, D. and {Krimm}, H.~A. and {Markwardt}, C.~B. and {McLean}, K. and {Palmer}, D. and {Parsons}, A. and {Tueller}, J.},
	doi = {10.1038/nature03515},
	eprint = {astro-ph/0503521},
	journal = {\nat},
	month = may,
	number = {7039},
	pages = {178-180},
	primaryclass = {astro-ph},
	title = {{A link between prompt optical and prompt {\ensuremath{\gamma}}-ray emission in {\ensuremath{\gamma}}-ray bursts}},
	volume = {435},
	year = 2005}

@article{1999Natur.398..400A,
	adsurl = {https://ui.adsabs.harvard.edu/abs/1999Natur.398..400A},
	archiveprefix = {arXiv},
	author = {{Akerlof}, C. and {Balsano}, R. and {Barthelmy}, S. and {Bloch}, J. and {Butterworth}, P. and {Casperson}, D. and {Cline}, T. and {Fletcher}, S. and {Frontera}, F. and {Gisler}, G. and {Heise}, J. and {Hills}, J. and {Kehoe}, R. and {Lee}, B. and {Marshall}, S. and {McKay}, T. and {Miller}, R. and {Piro}, L. and {Priedhorsky}, W. and {Szymanski}, J. and {Wren}, J.},
	doi = {10.1038/18837},
	eprint = {astro-ph/9903271},
	journal = {\nat},
	month = apr,
	number = {6726},
	pages = {400-402},
	primaryclass = {astro-ph},
	title = {{Observation of contemporaneous optical radiation from a {\ensuremath{\gamma}}-ray burst}},
	volume = {398},
	year = 1999}

@article{2017Natur.547..425T,
	adsurl = {https://ui.adsabs.harvard.edu/abs/2017Natur.547..425T},
	author = {{Troja}, E. and {Lipunov}, V.~M. and {Mundell}, C.~G. and {Butler}, N.~R. and {Watson}, A.~M. and {Kobayashi}, S. and {Cenko}, S.~B. and {Marshall}, F.~E. and {Ricci}, R. and {Fruchter}, A. and {Wieringa}, M.~H. and {Gorbovskoy}, E.~S. and {Kornilov}, V. and {Kutyrev}, A. and {Lee}, W.~H. and {Toy}, V. and {Tyurina}, N.~V. and {Budnev}, N.~M. and {Buckley}, D.~A.~H. and {Gonz{\'a}lez}, J. and {Gress}, O. and {Horesh}, A. and {Panasyuk}, M.~I. and {Prochaska}, J.~X. and {Ramirez-Ruiz}, E. and {Rebolo Lopez}, R. and {Richer}, M.~G. and {Roman-Zuniga}, C. and {Serra-Ricart}, M. and {Yurkov}, V. and {Gehrels}, N.},
	doi = {10.1038/nature23289},
	journal = {\nat},
	month = jul,
	pages = {425-427},
	title = {{Significant and variable linear polarization during the prompt optical flash of GRB 160625B.}},
	volume = {547},
	year = 2017}

@article{2023NatAs...7..843O,
	adsurl = {https://ui.adsabs.harvard.edu/abs/2023NatAs...7..843O},
	archiveprefix = {arXiv},
	author = {{Oganesyan}, Gor and {Karpov}, Sergey and {Salafia}, Om Sharan and {Jel{\'\i}nek}, Martin and {Beskin}, Gregory and {Ronchini}, Samuele and {Banerjee}, Biswajit and {Branchesi}, Marica and {{\v{S}}trobl}, Jan and {Pol{\'a}{\v{s}}ek}, Cyril and {Hudec}, Ren{\'e} and {Ivanov}, Eugeny and {Katkova}, Elena and {Perkov}, Alexey and {Biryukov}, Anton and {Lyapsina}, Nadezhda and {Sasyuk}, Vyacheslav and {Ma{\v{s}}ek}, Martin and {Jane{\v{c}}ek}, Petr and {Ebr}, Jan and {Jury{\v{s}}ek}, Jakub and {Cunniffe}, Ronan and {Prouza}, Michael},
	doi = {10.1038/s41550-023-01972-4},
	eprint = {2109.00010},
	journal = {Nature Astronomy},
	month = jul,
	pages = {843-855},
	primaryclass = {astro-ph.HE},
	title = {{Exceptionally bright optical emission from a rare and distant gamma-ray burst}},
	volume = {7},
	year = 2023}

@article{2021ApJ...922...78X,
	adsurl = {https://ui.adsabs.harvard.edu/abs/2021ApJ...922...78X},
	archiveprefix = {arXiv},
	author = {{Xin}, L.~P. and {Li}, H.~L. and {Wang}, J. and {Han}, X.~H. and {Qiu}, Y.~L. and {Cai}, H.~B. and {Niu}, C.~H. and {Lu}, X.~M. and {Liang}, E.~W. and {Dai}, Z.~G. and {Wang}, X.~G. and {Wang}, X.~Y. and {Huang}, L. and {Wu}, C. and {Li}, G.~W. and {Feng}, Q.~C. and {Deng}, J.~S. and {Sun}, S.~S. and {Yang}, Y.~G. and {Wei}, J.~Y.},
	doi = {10.3847/1538-4357/ac1daf},
	eid = {78},
	eprint = {2108.06931},
	journal = {\apj},
	month = nov,
	number = {1},
	pages = {78},
	primaryclass = {astro-ph.HE},
	title = {{Constraints on Optical Emission of FAST-detected FRB 20181130B with GWAC Synchronized Observations}},
	volume = {922},
	year = 2021}

@article{2002PASP..114..721P,
	adsurl = {https://ui.adsabs.harvard.edu/abs/2002PASP..114..721P},
	archiveprefix = {arXiv},
	author = {{Patterson}, Joseph and {Masi}, Gianluca and {Richmond}, Michael W. and {Martin}, Brian and {Beshore}, Edward and {Skillman}, David R. and {Kemp}, Jonathan and {Vanmunster}, Tonny and {Rea}, Robert and {Allen}, William and {Davis}, Stacey and {Davis}, Tracy and {Henden}, Arne A. and {Starkey}, Donn and {Foote}, Jerry and {Oksanen}, Arto and {Cook}, Lewis M. and {Fried}, Robert E. and {Husar}, Dieter and {Nov{\'a}k}, Rudolf and {Campbell}, Tut and {Robertson}, Jeff and {Krajci}, Thomas and {Pavlenko}, Elena and {Mirabal}, Nestor and {Niarchos}, Panos G. and {Brettman}, Orville and {Walker}, Stan},
	doi = {10.1086/341696},
	eprint = {astro-ph/0204126},
	journal = {\pasp},
	month = jul,
	number = {797},
	pages = {721-747},
	primaryclass = {astro-ph},
	title = {{The 2001 Superoutburst of WZ Sagittae}},
	volume = {114},
	year = 2002}

@article{2006PASJ...58L..23I,
	adsurl = {https://ui.adsabs.harvard.edu/abs/2006PASJ...58L..23I},
	archiveprefix = {arXiv},
	author = {{Imada}, Akira and {Kubota}, Kaori and {Kato}, Taichi and {Nogami}, Daisaku and {Maehara}, Hiroyuki and {Nakajima}, Kazuhiro and {Uemura}, Makoto and {Ishioka}, Ryoko},
	doi = {10.1093/pasj/58.4.L23},
	eprint = {astro-ph/0605286},
	journal = {\pasj},
	month = aug,
	pages = {L23-L27},
	primaryclass = {astro-ph},
	title = {{Discovery of a New Dwarf Nova, TSS J022216.4+412259.9: WZ Sge-Type Dwarf Nova Breaking the Shortest Superhump Period Record}},
	volume = {58},
	year = 2006}

@article{2015PASJ...67...52M,
	adsurl = {https://ui.adsabs.harvard.edu/abs/2015PASJ...67...52M},
	archiveprefix = {arXiv},
	author = {{Meyer}, Friedrich and {Meyer-Hofmeister}, Emmi},
	doi = {10.1093/pasj/psv023},
	eid = {52},
	eprint = {1504.00469},
	journal = {\pasj},
	month = jun,
	number = {3},
	pages = {52},
	primaryclass = {astro-ph.SR},
	title = {{SU UMa stars: Rebrightenings after superoutburst}},
	volume = {67},
	year = 2015}

@article{2018MNRAS.481.2523P,
	adsurl = {https://ui.adsabs.harvard.edu/abs/2018MNRAS.481.2523P},
	archiveprefix = {arXiv},
	author = {{Pala}, A.~F. and {Schmidtobreick}, L. and {Tappert}, C. and {G{\"a}nsicke}, B.~T. and {Mehner}, A.},
	doi = {10.1093/mnras/sty2434},
	eprint = {1809.02135},
	journal = {\mnras},
	month = dec,
	number = {2},
	pages = {2523-2535},
	primaryclass = {astro-ph.SR},
	title = {{The cataclysmic variable QZ Lib: a period bouncer}},
	volume = {481},
	year = 2018}

@book{1995cvs..book.....W,
	adsurl = {https://ui.adsabs.harvard.edu/abs/1995cvs..book.....W},
	author = {{Warner}, Brian},
	title = {{Cataclysmic variable stars}},
	volume = {28},
	year = 1995}

@article{1981A&A...104L..10M,
	adsurl = {https://ui.adsabs.harvard.edu/abs/1981A&A...104L..10M},
	author = {{Meyer}, F. and {Meyer-Hofmeister}, E.},
	journal = {\aap},
	month = jan,
	pages = {L10-L12},
	title = {{On the elusive cause of cataclysmic variable outbursts.}},
	volume = {104},
	year = 1981}

@article{1989PASJ...41.1005O,
	adsurl = {https://ui.adsabs.harvard.edu/abs/1989PASJ...41.1005O},
	author = {{Osaki}, Yoji},
	doi = {10.1093/pasj/41.5.1005},
	journal = {\pasj},
	month = dec,
	number = {5},
	pages = {1005-1033},
	title = {{A Model for the Superoutburst Phenomenon of SU Ursae Majoris Stars}},
	volume = {41},
	year = 1989}

@article{1996PASP..108...39O,
	adsurl = {https://ui.adsabs.harvard.edu/abs/1996PASP..108...39O},
	author = {{Osaki}, Yoji},
	doi = {10.1086/133689},
	journal = {\pasp},
	month = jan,
	pages = {39},
	title = {{Dwarf-Nova Outbursts}},
	volume = {108},
	year = 1996}

@article{2001NewAR..45..449L,
	adsurl = {https://ui.adsabs.harvard.edu/abs/2001NewAR..45..449L},
	archiveprefix = {arXiv},
	author = {{Lasota}, Jean-Pierre},
	doi = {10.1016/S1387-6473(01)00112-9},
	eprint = {astro-ph/0102072},
	journal = {\nar},
	month = jun,
	number = {7},
	pages = {449-508},
	primaryclass = {astro-ph},
	title = {{The disc instability model of dwarf novae and low-mass X-ray binary transients}},
	volume = {45},
	year = 2001}

@article{2017NatAs...1E.184S,
	adsurl = {https://ui.adsabs.harvard.edu/abs/2017NatAs...1E.184S},
	archiveprefix = {arXiv},
	author = {{Shulyak}, D. and {Reiners}, A. and {Engeln}, A. and {Malo}, L. and {Yadav}, R. and {Morin}, J. and {Kochukhov}, O.},
	doi = {10.1038/s41550-017-0184},
	eid = {0184},
	eprint = {1801.08571},
	journal = {Nature Astronomy},
	month = aug,
	pages = {0184},
	primaryclass = {astro-ph.SR},
	title = {{Strong dipole magnetic fields in fast rotating fully convective stars}},
	volume = {1},
	year = 2017}

@article{2022SerAJ.205....1L,
	adsurl = {https://ui.adsabs.harvard.edu/abs/2022SerAJ.205....1L},
	archiveprefix = {arXiv},
	author = {{Leitzinger}, M. and {Odert}, P.},
	doi = {10.2298/SAJ2205001L},
	eprint = {2212.09079},
	journal = {Serbian Astronomical Journal},
	month = dec,
	pages = {1-22},
	primaryclass = {astro-ph.SR},
	title = {{Stellar Coronal Mass Ejections}},
	volume = {205},
	year = 2022}

@article{2007AsBio...7..167K,
	adsurl = {https://ui.adsabs.harvard.edu/abs/2007AsBio...7..167K},
	author = {{Khodachenko}, Maxim L. and {Ribas}, Ignasi and {Lammer}, Helmut and {Grie{\ss}meier}, Jean-Mathias and {Leitner}, Martin and {Selsis}, Franck and {Eiroa}, Carlos and {Hanslmeier}, Arnold and {Biernat}, Helfried K. and {Farrugia}, Charles J. and {Rucker}, Helmut O.},
	doi = {10.1089/ast.2006.0127},
	journal = {Astrobiology},
	month = feb,
	number = {1},
	pages = {167-184},
	title = {{Coronal Mass Ejection (CME) Activity of Low Mass M Stars as An Important Factor for The Habitability of Terrestrial Exoplanets. I. CME Impact on Expected Magnetospheres of Earth-Like Exoplanets in Close-In Habitable Zones}},
	volume = {7},
	year = 2007}

@article{2016NatGe...9..452A,
	adsurl = {https://ui.adsabs.harvard.edu/abs/2016NatGe...9..452A},
	author = {{Airapetian}, V.~S. and {Glocer}, A. and {Gronoff}, G. and {H{\'e}brard}, E. and {Danchi}, W.},
	doi = {10.1038/ngeo2719},
	journal = {Nature Geoscience},
	month = jun,
	number = {6},
	pages = {452-455},
	title = {{Prebiotic chemistry and atmospheric warming of early Earth by an active young Sun}},
	volume = {9},
	year = 2016}

@article{2024ApJ...971..114L,
	adsurl = {https://ui.adsabs.harvard.edu/abs/2024ApJ...971..114L},
	archiveprefix = {arXiv},
	author = {{Li}, Guang-Wei and {Wang}, Liang and {Yuan}, Hai-Long and {Xin}, Li-Ping and {Wang}, Jing and {Wu}, Chao and {Li}, Hua-Li and {Haerken}, Hasitieer and {Wang}, Wei-Hua and {Cai}, Hong-Bo and {Han}, Xu-Hui and {Xu}, Yang and {Huang}, Lei and {Lu}, Xiao-Meng and {Bai}, Jian-Ying and {Wang}, Xiang-Yu and {Dai}, Zi-Gao and {Liang}, En-Wei and {Wei}, Jian-Yan},
	doi = {10.3847/1538-4357/ad55e8},
	eid = {114},
	eprint = {2407.08183},
	journal = {\apj},
	month = aug,
	number = {1},
	pages = {114},
	primaryclass = {astro-ph.SR},
	title = {{The White-light Superflares from Cool Stars in GWAC Triggers}},
	volume = {971},
	year = 2024}

@article{2023RAA....23a5016L,
	adsurl = {https://ui.adsabs.harvard.edu/abs/2023RAA....23a5016L},
	archiveprefix = {arXiv},
	author = {{Li}, Guang-Wei and {Wu}, Chao and {Zhou}, Gui-Ping and {Yang}, Chen and {Li}, Hua-Li and {Chen}, Jie and {Xin}, Li-Ping and {Wang}, Jing and {Haerken}, Hasitieer and {Ma}, Chao-Hong and {Cai}, Hong-Bo and {Han}, Xu-Hui and {Huang}, Lei and {Lu}, Xiao-Meng and {Bai}, Jian-Ying and {Zhang}, Xu-Kang and {Hao}, Xin-Li and {Wang}, Xiang-Yu and {Dai}, Zi-Gao and {Liang}, En-Wei and {Meng}, Xiao-Feng and {Wei}, Jian-Yan},
	doi = {10.1088/1674-4527/aca506},
	eid = {015016},
	eprint = {2211.11240},
	journal = {Research in Astronomy and Astrophysics},
	month = jan,
	number = {1},
	pages = {015016},
	primaryclass = {astro-ph.SR},
	title = {{Magnetic Activity and Parameters of 43 Flare Stars in the GWAC Archive}},
	volume = {23},
	year = 2023}

@article{2020RAA....20...13T,
	adsurl = {https://ui.adsabs.harvard.edu/abs/2020RAA....20...13T},
	archiveprefix = {arXiv},
	author = {{Turpin}, Damien and {Wu}, Chao and {Han}, Xu-Hui and {Xin}, Li-Ping and {Antier}, Sarah and {Leroy}, Nicolas and {Cao}, Li and {Cai}, Hong-Bo and {Cordier}, Bertrand and {Deng}, Jin-Song and {Dong}, Wen-Long and {Feng}, Qi-Chen and {Huang}, Lei and {Jia}, Lei and {Klotz}, Alain and {Lachaud}, Cyril and {Li}, Hua-Li and {Liang}, En-Wei and {Liu}, Shun-Fang and {Lu}, Xiao-Meng and {Meng}, Xian-Min and {Qiu}, Yu-Lei and {Wang}, Hui-Juan and {Wang}, Jing and {Wang}, Shen and {Wang}, Xiang-Gao and {Wei}, Jian-Yan and {Wu}, Bo-Bing and {Xiao}, Yu-Jie and {Xu}, Da-Wei and {Xu}, Yang and {Yang}, Yuan-Gui and {Zhang}, Pin-Pin and {Zhang}, Ruo-Song and {Zhang}, Shuang-Nan and {Zheng}, Ya-Tong and {Zou}, Si-Cheng},
	doi = {10.1088/1674-4527/20/1/13},
	eid = {013},
	eprint = {1902.08476},
	journal = {Research in Astronomy and Astrophysics},
	month = jan,
	number = {1},
	pages = {013},
	primaryclass = {astro-ph.IM},
	title = {{The mini-GWAC optical follow-up of gravitational wave alerts - results from the O2 campaign and prospects for the upcoming O3 run}},
	volume = {20},
	year = 2020}

@article{2020AJ....159...35W,
	adsurl = {https://ui.adsabs.harvard.edu/abs/2020AJ....159...35W},
	author = {{Wang}, J. and {Li}, H.~L. and {Xin}, L.~P. and {Han}, X.~H. and {Meng}, X.~M. and {Brink}, T.~G. and {Cai}, H.~B. and {Dai}, Z.~G. and {Filippenko}, A.~V. and {Hsia}, C.-H. and {Huang}, L. and {Jia}, L. and {Li}, G.~W. and {Li}, Y.~B. and {Liang}, E.~W. and {Lu}, X.~M. and {Mao}, J. and {Qiu}, P. and {Qiu}, Y.~L. and {Ren}, J.~J. and {Turpin}, D. and {Wang}, H.~J. and {Wang}, X.~G. and {Wang}, X.~Y. and {Wu}, C. and {Xu}, Y. and {Yan}, J.~Z. and {Zhang}, J.~B. and {Zheng}, W. and {Wei}, J.~Y.},
	doi = {10.3847/1538-3881/ab5855},
	eid = {35},
	journal = {\aj},
	month = feb,
	number = {2},
	pages = {35},
	title = {{Photometric and Spectroscopic Studies of Superoutbursts of Three Dwarf Novae Independently Identified by the SVOM/GWAC System in 2018}},
	volume = {159},
	year = 2020}

@article{2021ApJ...916...92W,
	adsurl = {https://ui.adsabs.harvard.edu/abs/2021ApJ...916...92W},
	archiveprefix = {arXiv},
	author = {{Wang}, J. and {Xin}, L.~P. and {Li}, H.~L. and {Li}, G.~W. and {Sun}, S.~S. and {Gao}, C. and {Han}, X.~H. and {Dai}, Z.~G. and {Liang}, E.~W. and {Wang}, X.~Y. and {Wei}, J.~Y.},
	doi = {10.3847/1538-4357/ac096f},
	eid = {92},
	eprint = {2106.04774},
	journal = {\apj},
	month = aug,
	number = {2},
	pages = {92},
	primaryclass = {astro-ph.SR},
	title = {{Detection of Flare-associated CME Candidates on Two M-dwarfs by GWAC and Fast, Time-resolved Spectroscopic Follow-ups}},
	volume = {916},
	year = 2021}

@article{2021ApJ...909..106X,
	adsurl = {https://ui.adsabs.harvard.edu/abs/2021ApJ...909..106X},
	archiveprefix = {arXiv},
	author = {{Xin}, L.~P. and {Li}, H.~L. and {Wang}, J. and {Han}, X.~H. and {Xu}, Y. and {Meng}, X.~M. and {Cai}, H.~B. and {Huang}, L. and {Lu}, X.~M. and {Qiu}, Y.~L. and {Wang}, X.~G. and {Liang}, E.~W. and {Dai}, Z.~G. and {Wang}, X.~Y. and {Wu}, C. and {Zhang}, J.~B. and {Li}, G.~W. and {Turpin}, D. and {Feng}, Q.~C. and {Deng}, J.~S. and {Sun}, S.~S. and {Zheng}, T.~C. and {Yang}, Y.~G. and {Wei}, J.~Y.},
	doi = {10.3847/1538-4357/abdd1b},
	eid = {106},
	eprint = {2012.14126},
	journal = {\apj},
	month = mar,
	number = {2},
	pages = {106},
	primaryclass = {astro-ph.SR},
	title = {{A {\ensuremath{\Delta}}R {\ensuremath{\sim}} 9.5 mag Superflare of an Ultracool Star Detected by the SVOM/GWAC System}},
	volume = {909},
	year = 2021}

@article{2016PASP..128k5005F,
	adsurl = {https://ui.adsabs.harvard.edu/abs/2016PASP..128k5005F},
	archiveprefix = {arXiv},
	author = {{Fan}, Zhou and {Wang}, Huijuan and {Jiang}, Xiaojun and {Wu}, Hong and {Li}, Hongbin and {Huang}, Yang and {Xu}, Dawei and {Hu}, Zhongwen and {Zhu}, Yinan and {Wang}, Jianfeng and {Komossa}, Stefanie and {Zhang}, Xiaoming},
	doi = {10.1088/1538-3873/128/969/115005},
	eprint = {1605.09166},
	journal = {\pasp},
	month = nov,
	number = {969},
	pages = {115005},
	primaryclass = {astro-ph.IM},
	title = {{The Xinglong 2.16-m Telescope: Current Instruments and Scientific Projects}},
	volume = {128},
	year = 2016}

@article{2017ApJ...846...31C,
	adsurl = {https://ui.adsabs.harvard.edu/abs/2017ApJ...846...31C},
	archiveprefix = {arXiv},
	author = {{Cherenkov}, A. and {Bisikalo}, D. and {Fossati}, L. and {M{\"o}stl}, C.},
	doi = {10.3847/1538-4357/aa82b2},
	eid = {31},
	eprint = {1709.01027},
	journal = {\apj},
	month = sep,
	number = {1},
	pages = {31},
	primaryclass = {astro-ph.EP},
	title = {{The Influence of Coronal Mass Ejections on the Mass-loss Rates of Hot-Jupiters}},
	volume = {846},
	year = 2017}

@article{2025GCN.42548....1L,
	adsurl = {https://ui.adsabs.harvard.edu/abs/2025GCN.42548....1L},
	author = {{Liang}, Y.~F. and {Zhou}, H. and {Pan}, H.~W. and {Einstein Probe Team}},
	journal = {GRB Coordinates Network},
	month = nov,
	pages = {1},
	title = {{EP251102a: Einstein Probe detection of a fast X-ray transient}},
	volume = {42548},
	year = 2025}

@article{2025GCN.42816....1Z,
	adsurl = {https://ui.adsabs.harvard.edu/abs/2025GCN.42816....1Z},
	author = {{Zhou}, H. and {Cao}, J.-Y. and {Yang}, H.-N. and {Ling}, Z.-X. and {Einstein Probe Team}},
	journal = {GRB Coordinates Network},
	month = nov,
	pages = {1},
	title = {{EP251124a: Einstein Probe detection of an X-ray transient}},
	volume = {42816},
	year = 2025}

@article{2025GCN.42800....1L,
	adsurl = {https://ui.adsabs.harvard.edu/abs/2025GCN.42800....1L},
	author = {{Lin}, Z.-Y. and {Lachaud}, C. and {Schanne}, S. and {Turpin}, D. and {Maiolino}, T. and {G{\"o}tz}, D. and {Malesani}, D.~B. and {SVOM Team}},
	journal = {GRB Coordinates Network},
	month = nov,
	pages = {1},
	title = {{GRB 251122A / EP251122a: SVOM detection of a long burst}},
	volume = {42800},
	year = 2025}

@article{2025GCN.42437....1H,
	adsurl = {https://ui.adsabs.harvard.edu/abs/2025GCN.42437....1H},
	author = {{Hussein}, S. and {Brunet}, M. and {Jacob}, U. and {Gotz}, D. and {Maggi}, P. and {SVOM mission Team}},
	journal = {GRB Coordinates Network},
	month = oct,
	pages = {1},
	title = {{GRB 251025B: SVOM detection of a burst}},
	volume = {42437},
	year = 2025}

@article{2025GCN.42006....1B,
	adsurl = {https://ui.adsabs.harvard.edu/abs/2025GCN.42006....1B},
	author = {{Barthelmy}, S.~D. and {Dichiara}, S. and {Gupta}, R. and {Krimm}, H.~A. and {Laha}, S. and {Lien}, A.~Y. and {Markwardt}, C.~B. and {Moss}, M.~J. and {Palmer}, D.~M. and {Parsotan}, T. and {Sadaula}, D. and {Sakamoto}, T.},
	journal = {GRB Coordinates Network},
	month = sep,
	pages = {1},
	title = {{GRB 250925A: Swift-BAT refined analysis}},
	volume = {42006},
	year = 2025}

@article{2025GCN.42001....1Z,
	adsurl = {https://ui.adsabs.harvard.edu/abs/2025GCN.42001....1Z},
	author = {{Zhou}, H. and {Hu}, D.~F. and {Yang}, Z.~H. and {Zhao}, Q.~C. and {Wang}, Y.~L. and {Jin}, C.~C. and {Einstein Probe Team}},
	journal = {GRB Coordinates Network},
	month = sep,
	pages = {1},
	title = {{GRB 250925A/EP250925a: EP-WXT detection}},
	volume = {42001},
	year = 2025}

@article{2025GCN.40050....1F,
	adsurl = {https://ui.adsabs.harvard.edu/abs/2025GCN.40050....1F},
	author = {{Fermi GBM Team}},
	journal = {GRB Coordinates Network},
	month = apr,
	pages = {1},
	title = {{GRB 250404A: Fermi GBM Final Real-time Localization}},
	volume = {40050},
	year = 2025}

@article{2025GCN.40051....1H,
	adsurl = {https://ui.adsabs.harvard.edu/abs/2025GCN.40051....1H},
	author = {{Hu}, J.~W. and {Liu}, Q.~C. and {Zhang}, B.~B. and {Hua}, Y.~L. and {Liu}, Y. and {Einstein Probe Team}},
	journal = {GRB Coordinates Network},
	month = apr,
	pages = {1},
	title = {{EP250404a: Einstein Probe detection of an X-ray transient}},
	volume = {40051},
	year = 2025}

@inproceedings{2020SPIE11445E..7MY,
	adsurl = {https://ui.adsabs.harvard.edu/abs/2020SPIE11445E..7MY},
	author = {{Yuan}, Xiangyan and {Li}, Zhengyang and {Liu}, Xiaowei and {Niu}, Dongsheng and {Lu}, Qishui and {Jiang}, Fanghua and {Wang}, Yuefei and {Li}, Xiaoyan and {Liang}, YongJun and {Wang}, Hai and {Zhang}, Chao and {Wang}, Jinfeng and {Li}, Bo and {Tian}, Jie and {Lu}, Haiping and {Chen}, Bingqiu and {Huang}, Yang and {Liu}, Xiangkun and {Yao}, Zhengqiu and {Cui}, Xiangqun and {Li}, Guoping},
	booktitle = {Ground-based and Airborne Telescopes VIII},
	doi = {10.1117/12.2562334},
	editor = {{Marshall}, Heather K. and {Spyromilio}, Jason and {Usuda}, Tomonori},
	eid = {114457M},
	month = dec,
	pages = {114457M},
	series = {Society of Photo-Optical Instrumentation Engineers (SPIE) Conference Series},
	title = {{Development of the Multi-channel Photometric Survey telescope}},
	volume = {11445},
	year = 2020}

@article{2023SCPMA..6609512W,
	adsurl = {https://ui.adsabs.harvard.edu/abs/2023SCPMA..6609512W},
	archiveprefix = {arXiv},
	author = {{Wang}, Tinggui and {Liu}, Guilin and {Cai}, Zhenyi and {Geng}, Jinjun and {Fang}, Min and {He}, Haoning and {Jiang}, Ji-an and {Jiang}, Ning and {Kong}, Xu and {Li}, Bin and {Li}, Ye and {Luo}, Wentao and {Pan}, Zhizheng and {Wu}, Xuefeng and {Yang}, Ji and {Yu}, Jiming and {Zheng}, Xianzhong and {Zhu}, Qingfeng and {Cai}, Yi-Fu and {Chen}, Yuanyuan and {Chen}, Zhiwei and {Dai}, Zigao and {Fan}, Lulu and {Fan}, Yizhong and {Fang}, Wenjuan and {He}, Zhicheng and {Hu}, Lei and {Hu}, Maokai and {Jin}, Zhiping and {Jiang}, Zhibo and {Li}, Guoliang and {Li}, Fan and {Li}, Xuzhi and {Liang}, Runduo and {Lin}, Zheyu and {Liu}, Qingzhong and {Liu}, Wenhao and {Liu}, Zhengyan and {Liu}, Wei and {Liu}, Yao and {Lou}, Zheng and {Qu}, Han and {Sheng}, Zhenfeng and {Shi}, Jianchun and {Shu}, Yiping and {Su}, Zhenbo and {Sun}, Tianrui and {Wang}, Hongchi and {Wang}, Huiyuan and {Wang}, Jian and {Wang}, Junxian and {Wei}, Daming and {Wei}, Junjie and {Xue}, Yongquan and {Yan}, Jingzhi and {Yang}, Chao and {Yuan}, Ye and {Yuan}, Yefei and {Zhang}, Hongxin and {Zhang}, Miaomiao and {Zhao}, Haibin and {Zhao}, Wen},
	doi = {10.1007/s11433-023-2197-5},
	eid = {109512},
	eprint = {2306.07590},
	journal = {Science China Physics, Mechanics, and Astronomy},
	month = oct,
	number = {10},
	pages = {109512},
	primaryclass = {astro-ph.IM},
	title = {{Science with the 2.5-meter Wide Field Survey Telescope (WFST)}},
	volume = {66},
	year = 2023}

@article{2025RAA....25d4001H,
	adsurl = {https://ui.adsabs.harvard.edu/abs/2025RAA....25d4001H},
	archiveprefix = {arXiv},
	author = {{Huang}, Yang and {Liu}, Jifeng and {Wu}, Hong and {Shang}, Zhaohui and {Luo}, Ali and {Hu}, Shaoming and {Cui}, Wenyuan and {Mao}, Yongna},
	doi = {10.1088/1674-4527/adc795},
	eid = {044001},
	eprint = {2504.01615},
	journal = {Research in Astronomy and Astrophysics},
	month = apr,
	number = {4},
	pages = {044001},
	primaryclass = {astro-ph.IM},
	title = {{The Mini-SiTian Array: A Pathfinder for the SiTian Project}},
	volume = {25},
	year = 2025}

@article{2023ApJS..265...63C,
	adsurl = {https://ui.adsabs.harvard.edu/abs/2023ApJS..265...63C},
	archiveprefix = {arXiv},
	author = {{Corbett}, Hank and {Carney}, Jonathan and {Gonzalez}, Ramses and {Fors}, Octavi and {Galliher}, Nathan and {Glazier}, Amy and {Howard}, Ward S. and {Law}, Nicholas M. and {Quimby}, Robert and {Ratzloff}, Jeffrey K. and {Soto}, Alan Vasquez},
	doi = {10.3847/1538-4365/acbd41},
	eid = {63},
	eprint = {2302.10929},
	journal = {\apjs},
	month = apr,
	number = {2},
	pages = {63},
	primaryclass = {astro-ph.IM},
	title = {{The Evryscope Fast Transient Engine: Real-time Detection for Rapidly Evolving Transients}},
	volume = {265},
	year = 2023}

@article{Han+etal+2026,
	author = {{Han}, X. H. and {Zhan}, P. P. and {Xiao}, Y. J. and {Xin}, L. P. and {Zhang}, R. S. and {Huang}, L. and {Lu}, X. M.},
	journal = {\raa},
	pages = {1-10},
	title = {{Alert Chain and Observation Planning for GWAC-N}},
	volume = {this issue},
	year = 2026}

@article{2019GCN.25606....1L,
	adsurl = {https://ui.adsabs.harvard.edu/abs/2019GCN.25606....1L},
	author = {{LIGO Scientific Collaboration} and {Virgo Collaboration}},
	journal = {GRB Coordinates Network},
	month = sep,
	pages = {1},
	title = {{LIGO/Virgo S190901ap: Identification of a GW compact binary merger candidate}},
	volume = {25606},
	year = 2019}

@article{2023PASP..135f5001O,
	adsurl = {https://ui.adsabs.harvard.edu/abs/2023PASP..135f5001O},
	archiveprefix = {arXiv},
	author = {{Ofek}, E.~O. and {Ben-Ami}, S. and {Polishook}, D. and {Segre}, E. and {Blumenzweig}, A. and {Strotjohann}, N.-L. and {Yaron}, O. and {Shani}, Y.~M. and {Nachshon}, S. and {Shvartzvald}, Y. and {Hershko}, O. and {Engel}, M. and {Segre}, M. and {Segev}, N. and {Zimmerman}, E. and {Nir}, G. and {Judkovsky}, Y. and {Gal-Yam}, A. and {Zackay}, B. and {Waxman}, E. and {Kushnir}, D. and {Chen}, P. and {Azaria}, R. and {Manulis}, I. and {Diner}, O. and {Vandeventer}, B. and {Franckowiak}, A. and {Weimann}, S. and {Borowska}, J. and {Garrappa}, S. and {Zenin}, A. and {Fallah Ramazani}, V. and {Konno}, R. and {K{\"u}sters}, D. and {Sadeh}, I. and {Parsons}, R.~D. and {Berge}, D. and {Kowalski}, M. and {Ohm}, S. and {Arcavi}, I. and {Bruch}, R.},
	doi = {10.1088/1538-3873/acd8f0},
	eid = {065001},
	eprint = {2304.04796},
	journal = {\pasp},
	month = jun,
	number = {1048},
	pages = {065001},
	primaryclass = {astro-ph.IM},
	title = {{The Large Array Survey Telescope-System Overview and Performances}},
	volume = {135},
	year = 2023}

@article{2019GCN.24168....1L,
	adsurl = {https://ui.adsabs.harvard.edu/abs/2019GCN.24168....1L},
	author = {{Ligo Scientific Collaboration} and {VIRGO Collaboration}},
	journal = {GRB Coordinates Network},
	month = jan,
	pages = {1},
	title = {{LIGO/Virgo S190425z: Identification of a GW compact binary merger candidate.}},
	volume = {24168},
	year = 2019}

@article{Cordier+etal+2026a,
	author = {{Cordier}, B. and {Wei}, J.Y. and {Zhang}, S.N. and {Basa}, S. and {Atteia}, J.-L. and {Others}},
	journal = {\raa},
	pages = {1-10},
	title = {{The SVOM mission, its profile and its system}},
	volume = {this issue},
	year = 2026}

@article{2009ApJ...702..791M,
	adsurl = {https://ui.adsabs.harvard.edu/abs/2009ApJ...702..791M},
	archiveprefix = {arXiv},
	author = {{Meegan}, Charles and {Lichti}, Giselher and {Bhat}, P.~N. and {Bissaldi}, Elisabetta and {Briggs}, Michael S. and {Connaughton}, Valerie and {Diehl}, Roland and {Fishman}, Gerald and {Greiner}, Jochen and {Hoover}, Andrew S. and {van der Horst}, Alexander J. and {von Kienlin}, Andreas and {Kippen}, R. Marc and {Kouveliotou}, Chryssa and {McBreen}, Sheila and {Paciesas}, W.~S. and {Preece}, Robert and {Steinle}, Helmut and {Wallace}, Mark S. and {Wilson}, Robert B. and {Wilson-Hodge}, Colleen},
	doi = {10.1088/0004-637X/702/1/791},
	eprint = {0908.0450},
	journal = {\apj},
	month = sep,
	number = {1},
	pages = {791-804},
	primaryclass = {astro-ph.IM},
	title = {{The Fermi Gamma-ray Burst Monitor}},
	volume = {702},
	year = 2009}

@incollection{2022hxga.book...86Y,
	adsurl = {https://ui.adsabs.harvard.edu/abs/2022hxga.book...86Y},
	author = {{Yuan}, Weimin and {Zhang}, Chen and {Chen}, Yong and {Ling}, Zhixing},
	booktitle = {Handbook of X-ray and Gamma-ray Astrophysics},
	doi = {10.1007/978-981-16-4544-0_151-1},
	editor = {{Bambi}, Cosimo and {Sangangelo}, Andrea},
	eid = {86},
	pages = {86},
	title = {{The Einstein Probe Mission}},
	year = 2022}

@article{2015PASJ...67..108K,
	adsurl = {https://ui.adsabs.harvard.edu/abs/2015PASJ...67..108K},
	archiveprefix = {arXiv},
	author = {{Kato}, Taichi},
	doi = {10.1093/pasj/psv077},
	eid = {108},
	eprint = {1507.07659},
	journal = {\pasj},
	month = dec,
	number = {6},
	pages = {108},
	primaryclass = {astro-ph.SR},
	title = {{WZ Sge-type dwarf novae}},
	volume = {67},
	year = 2015}

@article{2021NatAs...5..378L,
	adsurl = {https://ui.adsabs.harvard.edu/abs/2021NatAs...5..378L},
	archiveprefix = {arXiv},
	author = {{Li}, C.~K. and {Lin}, L. and {Xiong}, S.~L. and {Ge}, M.~Y. and {Li}, X.~B. and {Li}, T.~P. and {Lu}, F.~J. and {Zhang}, S.~N. and {Tuo}, Y.~L. and {Nang}, Y. and {Zhang}, B. and {Xiao}, S. and {Chen}, Y. and {Song}, L.~M. and {Xu}, Y.~P. and {Liu}, C.~Z. and {Jia}, S.~M. and {Cao}, X.~L. and {Qu}, J.~L. and {Zhang}, S. and {Gu}, Y.~D. and {Liao}, J.~Y. and {Zhao}, X.~F. and {Tan}, Y. and {Nie}, J.~Y. and {Zhao}, H.~S. and {Zheng}, S.~J. and {Zheng}, Y.~G. and {Luo}, Q. and {Cai}, C. and {Li}, B. and {Xue}, W.~C. and {Bu}, Q.~C. and {Chang}, Z. and {Chen}, G. and {Chen}, L. and {Chen}, T.~X. and {Chen}, Y.~B. and {Chen}, Y.~P. and {Cui}, W. and {Cui}, W.~W. and {Deng}, J.~K. and {Dong}, Y.~W. and {Du}, Y.~Y. and {Fu}, M.~X. and {Gao}, G.~H. and {Gao}, H. and {Gao}, M. and {Gu}, Y.~D. and {Guan}, J. and {Guo}, C.~C. and {Han}, D.~W. and {Huang}, Y. and {Huo}, J. and {Jiang}, L.~H. and {Jiang}, W.~C. and {Jin}, J. and {Jin}, Y.~J. and {Kong}, L.~D. and {Li}, G. and {Li}, M.~S. and {Li}, W. and {Li}, X. and {Li}, X.~F. and {Li}, Y.~G. and {Li}, Z.~W. and {Liang}, X.~H. and {Liu}, B.~S. and {Liu}, G.~Q. and {Liu}, H.~W. and {Liu}, X.~J. and {Liu}, Y.~N. and {Lu}, B. and {Lu}, X.~F. and {Luo}, T. and {Ma}, X. and {Meng}, B. and {Ou}, G. and {Sai}, N. and {Shang}, R.~C. and {Song}, X.~Y. and {Sun}, L. and {Tao}, L. and {Wang}, C. and {Wang}, G.~F. and {Wang}, J. and {Wang}, W.~S. and {Wang}, Y.~S. and {Wen}, X.~Y. and {Wu}, B.~B. and {Wu}, B.~Y. and {Wu}, M. and {Xiao}, G.~C. and {Xu}, H. and {Yang}, J.~W. and {Yang}, S. and {Yang}, Y.~J. and {Yang}, Yi-Jung and {Yi}, Q.~B. and {Yin}, Q.~Q. and {You}, Y. and {Zhang}, A.~M. and {Zhang}, C.~M. and {Zhang}, F. and {Zhang}, H.~M. and {Zhang}, J. and {Zhang}, T. and {Zhang}, W. and {Zhang}, W.~C. and {Zhang}, W.~Z. and {Zhang}, Y. and {Zhang}, Yue and {Zhang}, Y.~F. and {Zhang}, Y.~J. and {Zhang}, Z. and {Zhang}, Zhi and {Zhang}, Z.~L. and {Zhou}, D.~K. and {Zhou}, J.~F. and {Zhu}, Y. and {Zhu}, Y.~X. and {Zhuang}, R.~L.},
	doi = {10.1038/s41550-021-01302-6},
	eprint = {2005.11071},
	journal = {Nature Astronomy},
	month = apr,
	pages = {378-384},
	primaryclass = {astro-ph.HE},
	title = {{HXMT identification of a non-thermal X-ray burst from SGR J1935+2154 and with FRB 200428}},
	volume = {5},
	year = 2021}

\clearpage

\end{CJK*}
\end{document}